\documentclass[english,twocolumn,prl,amsmath,amssymb,floatfix,superscriptaddress]{revtex4-1}
\pdfpageattr {/Group << /S /Transparency /I true /CS /DeviceRGB>>}
\usepackage[T1]{fontenc}
\usepackage[latin9]{inputenc}
\usepackage{dcolumn}
\usepackage{bm}
\usepackage{graphicx}
\usepackage{color}
\usepackage{babel}
\usepackage{amsfonts}
\usepackage{mathtools}
\usepackage{slashed}
\usepackage{enumerate}
\usepackage{braket}
\usepackage[multiple]{footmisc}

\usepackage{tikz}
\usetikzlibrary{calc,shapes.arrows,shapes.geometric}
\begin{document}


\title{SO(5) non-Fermi liquid in a Coulomb box device}

\author{Andrew K. Mitchell}
\affiliation{School of Physics, University College Dublin, Belfield, Dublin 4, Ireland}
\affiliation{Centre for Quantum Engineering, Science, and Technology, University College Dublin, Belfield, Dublin 4, Ireland}
\author{Alon Liberman}
\author{Eran Sela}
\affiliation{School of Physics and Astronomy,
Tel Aviv University, Tel Aviv 6997801, Israel}
\author{Ian Affleck}
\affiliation{Department of Physics and Astronomy and Stewart Blusson Quantum Matter Institute, 
University of British Columbia, Vancouver, B.C., Canada, V6T 1Z1}


\begin{abstract}
\noindent Non-Fermi liquid (NFL) physics can be realized in quantum dot devices where competing interactions frustrate the exact screening of dot spin or charge degrees of freedom. 
We show that a standard nanodevice architecture, involving a dot coupled to both a quantum box and metallic leads, can host an exotic SO(5) symmetry Kondo effect, with entangled dot and box charge and spin. 
This NFL state is surprisingly robust to breaking channel and spin symmetry, but destabilized by particle-hole asymmetry. 
By tuning gate voltages, the SO(5) state evolves continuously to a spin and then ``flavor'' two-channel Kondo state. 
The expected experimental conductance signatures are highlighted. 
\end{abstract}
\maketitle


Nanoelectronic circuit realizations of fundamental quantum impurity models allow the nontrivial physics associated with strong electron correlations to be probed via quantum transport measurements \cite{sohn2013mesoscopic}. Quantum dot devices, in particular, can exhibit the Kondo effect at low temperatures \cite{pustilnik2004kondo}: a localized magnetic moment on the dot is dynamically screened by conduction electrons in the metallic leads. Single-dot devices can behave as single-electron transistors, with Kondo-enhanced spin-flip scattering strongly boosting the conductance between source and drain leads measured in experiments \cite{goldhaber1998kondo,cronenwett1998tunable,van2000kondo}.

The conventional Kondo effect \cite{hewson1997kondo} involves a localized ``impurity'' spin-$\tfrac{1}{2}$ degree of freedom, coupled to a single effective channel of conduction electrons, and has SU(2) spin symmetry. However, the Kondo effect is also observed in more complex systems, such as coupled quantum dot devices \cite{jeong2001kondo,malecki2010prospect} and single-molecule transistors \cite{liang2002kondo,mitchell2017kondo}, involving  spin and orbital degrees of freedom. In such systems, it is possible to realize variants of the classic spin-$\tfrac{1}{2}$ single-channel Kondo paradigm; e.g.~orbital \cite{lopez2005probing}, spin-1  \cite{sasaki2000kondo,paaske2006non}, and ferromagnetic \cite{mitchell2009quantum} Kondo effects. In particular, the symmetry of the effective model is important in determining the low-energy physics. Kondo effects with SU(4) symmetry can be realized in double quantum dots \cite{keller2014emergent,borda20034,*galpin2005quantum} and carbon nanotube dots \cite{choi20054,anders2008zero}, and also have Fermi liquid (FL) ground states.

More exotic non-Fermi liquid (NFL) states can be realized in multi-channel systems, where competing interactions frustrate exact screening of the dot spin or charge degrees of freedom at special high-symmetry points \cite{nozieres1980kondo}. This results in a residual dot entropy characteristic of fractionalized excitations, and anomalous conductance signatures \cite{affleck1991universal,affleck1993exact}. However, this kind of NFL physics is typically delicate, being found at the quantum critical point between more standard FL phases, and is unstable to relevant symmetry-breaking perturbations.

Experimentally, the major challenge to realize NFL Kondo physics in quantum dot devices is to prevent mixing between multiple conduction electron channels. Two prominent scenarios to achieve this utilize an interacting quantum box (``Coulomb box'') \cite{oreg2003two,furusaki1995theory}. The quantum box is a large quantum dot, hosting a macroscopically large number of electrons, but due to quantum confinement has a discrete level spacing $\delta$ and finite charging energy $E_C$. For $\delta < T < E_C$ the box effectively provides a continuum reservoir of conduction electrons, but also displays charge quantization \cite{matveev1995coulomb}. 

Spin-two channel Kondo (s-2CK) physics can be realized in a device involving a small quantum dot coupled to a quantum box as well as metallic leads \cite{oreg2003two}. 
The low-energy effective model consists of a dot spin-$\tfrac{1}{2}$ exchange coupled to two conduction electron channels (leads and box), with mixing between the channels suppressed by the large box charging energy. Both channels compete to Kondo-screen the dot spin, resulting in an NFL state. Breaking channel or spin symmetry relieves the frustration and results in a standard FL state. This physics was realized experimentally in Refs.~\cite{potok2007observation,keller2015universal}.

\begin{figure}[b!]
\includegraphics[width=8.7cm]{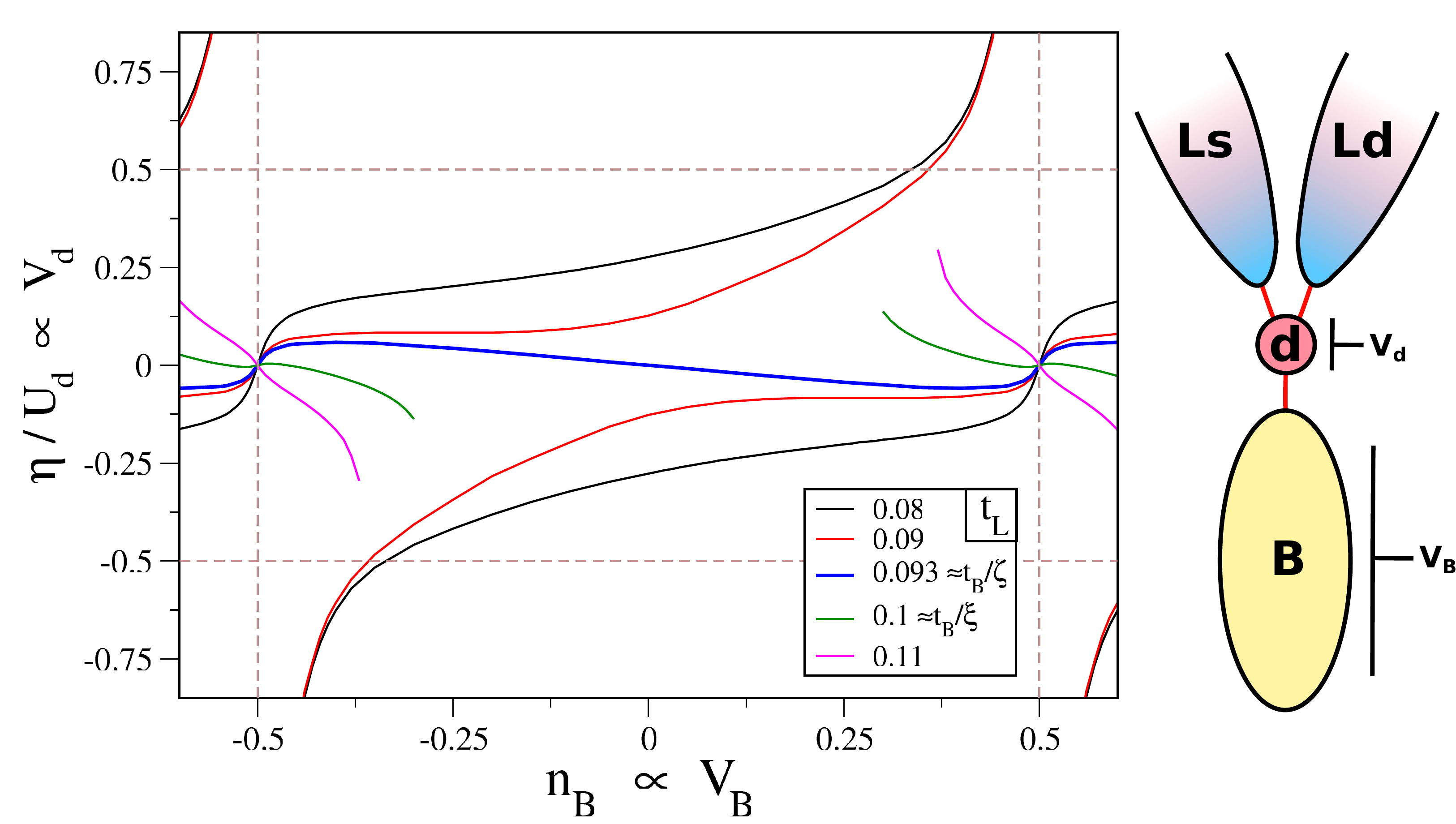}
  \caption{\textit{Right:} Schematic of the device: a quantum dot coupled to a quantum box and source/drain leads. \textit{Left:} NRG phase diagram spanned by dot and box gate voltages, $V_{\rm d}\propto \eta/U_d$ and $V_{\rm B}\propto n_{\rm B}$, showing the NFL line for various channel asymmetries $t_{\rm L}/t_{\rm B}$. SO(5) point located at $n_{\rm B}=\pm\tfrac{1}{2}$ and $\eta=0$. Plotted for constant $U_{d}=0.3$, $E_C=0.1$ and $t_{\rm B}=0.12$.}
  \label{fig:phase_diagram}
\end{figure}


By contrast, a charge-2CK (c-2CK) effect can be realized when a quantum box tuned to its charge degeneracy point is coupled to two leads, as proposed in Ref.~\cite{furusaki1995theory} and realized experimentally in Refs.~\cite{iftikhar2015two,iftikhar2018tunable}. In this case, the macroscopic box charge states play the role of a pseudospin impurity. Distinctive signatures of the resulting NFL state are observable in quantum transport \cite{iftikhar2015two,iftikhar2018tunable,mitchell2016universality,landau2018charge,nguyen2020thermoelectric,karki2020quantum,van2020wiedemann,*van2020electric,lee2020fractional}.

In this Letter, we revisit the device of Refs.~\cite{oreg2003two,potok2007observation,keller2015universal} but now examine the full phase diagram as function of dot and box gate voltages, which in turn control the dot and box occupancies -- see Fig.~\ref{fig:phase_diagram}. 
We show that the emergent SU(4) symmetry of the system arising when the dot hosts a local moment and the box is at its charge degeneracy point, is reduced to SO(5) at particle-hole symmetry. Although the SU(4) state is an FL \cite{le2003smearing}, a novel NFL Kondo effect arises at the SO(5) point, in which both dot and box charge and spin are maximally entangled. We achieve a detailed understanding of this state using a combination of conformal field theory \cite{affleck1990current,affleck1991critical,affleck1995conformal} (CFT), bosonization \cite{emery1992mapping}, and numerical renormalization group \cite{wilson1975renormalization,*bulla2008numerical,mitchell2014generalized,*stadler2016interleaved} (NRG) techniques. Remarkably, the NFL physics at this point is robust to breaking channel and/or spin symmetry. Furthermore, we show that by tuning gate voltages, the SO(5) state evolves continuously into the more familiar s-2CK state of Refs.~\cite{oreg2003two,potok2007observation,keller2015universal}, and then into a ``flavor''-2CK (f-2CK) effect when the dot local moment is lost but box charge fluctuations persist. The distinctive transport signatures associated with this physics are accessible in existing experimental setups.

\textit{Models and mappings.--} 
The device illustrated in Fig.~\ref{fig:phase_diagram} is described by the Hamiltonian $H=H_0+H_{\rm B}+H_{\rm d}+\sum_{\gamma}H_{\rm hyb}^{\gamma}$, with $\gamma=\rm Ls,~Ld,~B$ for the source/drain leads and box, respectively. $H_0=\sum_{\gamma,k,\sigma}\epsilon_{\gamma k}^{\phantom{\dagger}}  c_{\gamma k\sigma}^{\dagger}c_{\gamma k\sigma}^{\phantom{\dagger}}$ describes the three conduction electron reservoirs, while  
\begin{align}
H_{\rm B} &= E_C \left ( \hat{N}_{\rm B} - N_0 - n_{\rm B} \right )^2 \;, \label{eq:Hb} \\
H_{\rm d} &= \sum_{\sigma} \epsilon_d d_{\sigma}^{\dagger} d_{\sigma}^{\phantom{\dagger}} + U_d d_{\uparrow}^{\dagger} d_{\uparrow}^{\phantom{\dagger}} d_{\downarrow}^{\dagger} d_{\downarrow}^{\phantom{\dagger}} \;, \label{eq:Hd}
\end{align}
describe the box Coulomb interaction and the dot. The dot is tunnel-coupled to the leads and box via $H_{\rm hyb}^{\gamma}=\sum_{k,\sigma}(t_{\gamma k}d_{\sigma}^{\dagger}c_{\gamma k\sigma}^{\phantom{\dagger}} + \rm H.c.)$. 
Here, $\sigma =\uparrow,\downarrow$ denotes (real) spin, and $d_{\sigma}$ or $c_{\gamma k \sigma}$ are operators for the dot or conduction electrons, respectively. 
$\hat{N}_{\rm B}=\sum_{k,\sigma} c_{{\rm B} k\sigma}^{\dagger}c_{{\rm B} k\sigma}^{\phantom{\dagger}}$ is the total number operator for the box electrons. The dot and box occupations are controllable by gate voltages $V_{\rm d} \propto \eta =\epsilon_d+\tfrac{1}{2}U_d$ and $V_{\rm B} \propto n_{\rm B}$, respectively.  
For simplicity we now take equivalent conduction electron baths
$\epsilon_{\gamma k}\equiv \epsilon_k$ with a constant density of states $\nu$ defined inside a band of halfwidth $D=1$, such that $\epsilon_k = v_F k$ at low energies. We define $t_{\gamma}^2 = \sum_k |t_{\gamma k}|^2$ and $t_{\rm L}^2=t_{\rm Ls}^2+ t_{\rm Ld}^2$.

Following Ref.~\cite{anders2004coulomb}, we incorporate the box interaction term, Eq.~\ref{eq:Hb}, into the hybridization,
\begin{equation}
\label{eq:ALS}
H_{\rm B}+H_{\rm hyb}^{\rm B} ~\rightarrow~ E_C\left ( \hat{\mathcal{T}}^z-n_{\rm B} \right )^2 + \sum_{k,\sigma}(t_{{\rm B} k}d_{\sigma}^{\dagger}c_{{\rm B} k\sigma}^{\phantom{\dagger}}\hat{\mathcal{T}}^{-} + \rm H.c.) \;,\nonumber
\end{equation}
where $\hat{\mathcal{T}}^{\pm}=\sum_{N_{\rm B}} |N_{\rm B}\pm 1\rangle\langle N_{\rm B}|$ are ladder operators for the box charge, and $\hat{\mathcal{T}}^z=\sum_{N_{\rm B}}(N_{\rm B}-N_0)|N_{\rm B}\rangle\langle N_{\rm B}|$. 
Note that the model possesses the symmetry $n_{\rm B}\rightarrow n_{\rm B}\pm 1$. Particle-hole (ph) asymmetry is controlled by $n_{\rm B}$ and $\eta$; the model is invariant to replacing $n_{\rm B}\rightarrow -n_{\rm B}$ and $\eta \rightarrow -\eta$, related by a ph transformation. Exact ph symmetry arises at $\eta=0$ for any integer or half-integer $n_{\rm B}$.

For the NRG calculations presented here, only a finite number of charge states around the reference $N_0$ are required to obtain converged results \cite{wilson1975renormalization,*bulla2008numerical,mitchell2014generalized,*stadler2016interleaved,SM}.


\textit{Spin-2CK regime.--} 
For large box charging energy $E_C$ and deep in the dot and box Coulomb blockade regime (near the point $\eta=0$ and $n_{\rm B} = 0$), the dot hosts an effective spin-$\tfrac{1}{2}$ local moment, and the box has a well-defined number of electrons $N_0$. At low temperatures $T\ll E_C,U_d$ virtual charge fluctuations on the dot and box due to $H_{\rm hyb}$ generate the spin-flip scattering responsible for the Kondo effect. However, finite $E_C$ blocks charge transfer between the leads and box, giving rise to a frustration of Kondo screening and the possibility of NFL physics \cite{oreg2003two}. In this regime, a standard Schrieffer-Wolff transformation (SWT) yields the s-2CK model \cite{oreg2003two,nozieres1980kondo},
\begin{align}
\label{eq:H2ck}
H_{s-2CK} = H_0 + \vec{S}_d \cdot \left(\mathcal{J}_{\rm L} \vec{S}_{\rm L} + \mathcal{J}_{\rm B}\vec{S}_{\rm B}\right ) \;,
\end{align}
where $\vec{S}_d$ is a spin-$\tfrac{1}{2}$ operator for the dot, while $\vec{S}_{\alpha=\rm L,B} = \tfrac{1}{2}\sum_{\sigma,\sigma'}c_{\alpha  \sigma}^{\dagger} \vec{\boldsymbol{\sigma}}_{\sigma\sigma'} c_{\alpha \sigma'}$, with $c_{{\rm B}\sigma} =\tfrac{1}{t_{\rm B}}\sum_k t_{{\rm B} k} c_{{\rm B} k \sigma}$ and $c_{{\rm L}\sigma} =\tfrac{1}{t_{\rm L}}\sum_k( t_{{\rm Ls} k} c_{{\rm Ls} k \sigma}+t_{{\rm Ld} k} c_{{\rm Ld} k \sigma} )$ the local conduction electron orbitals at the dot position, and where
\begin{align}
\label{eq:s2ckJ}
\mathcal{J}_{\rm L}=\frac{8t_{\rm L}^2}{U_d}\left[1-\left(\tfrac{2\eta}{U_d}\right)^2\right]^{-1} \;\; ; \;\;
\mathcal{J}_{\rm B}=\frac{8t_{\rm B}^2}{U'_d}\left[1-\left(\tfrac{2\eta'}{U'_d}\right)^2\right]^{-1}
\end{align}
with $U'_d=U_d+2E_C$ and $\eta'=\eta+2E_C n_{\rm B}$.
Deep in the s-2CK regime, NFL physics arises  when $\mathcal{J}_{\rm L}=\mathcal{J}_{\rm B}$. For given physical device parameters $U_d, E_C, t_{\rm L}, t_{\rm B}$, Eq.~\ref{eq:s2ckJ} implies the existence of two NFL \textit{lines} in the $(n_{\rm B},\eta)$ plane related by the symmetry $\eta\rightarrow -\eta$ and $n_{\rm B}\rightarrow -n_{\rm B}$, see Fig.~\ref{fig:phase_diagram}. NFL physics can therefore be accessed by tuning the gate voltages $V_{\rm d}\propto \eta$ and $V_{\rm B}\propto n_{\rm B}$, as demonstrated experimentally in this regime in Refs.~\cite{potok2007observation,keller2015universal}. 
At the ph symmetric point $\eta=n_{\rm B}=0$, s-2CK arises for $t_{\rm B}=\zeta t_{\rm L}$ with $\zeta^2\simeq 1+2E_C/U_d$. Although this NFL state is robust to ph asymmetry, it is destabilized by channel asymmetry $\mathcal{J}_{\rm L}\ne \mathcal{J}_{\rm B}$ or spin asymmetry $B\ne 0$~\cite{affleck1992relevance}.


\textit{SO(5) Kondo.--} 
At $n_{\rm B}=\tfrac{1}{2}$, the box states with $N_0$ and $(N_0+1)$ electrons are exactly degenerate. Neglecting other box charge states (which are $\sim E_C$ higher in energy), we may define charge pseudospin-$\tfrac{1}{2}$ operators $\hat{T}_{\rm B}^{+}=|N_0+1\rangle\langle N_0|$, $\hat{T}_{\rm B}^{-}=(\hat{T}_{\rm B}^+)^{\dagger}$, and $\hat{T}_{\rm B}^z=\tfrac{1}{2}(|N_0+1\rangle\langle N_0+1| - |N_0\rangle\langle N_0|)$. The charge pseudospin is flipped by electronic tunneling between the dot and box. The low-energy effective model is obtained by projecting onto the dot spin and box pseudospin sectors using a generalized SWT. We now consider explicitly the special point with ph symmetry ($n_{\rm B}=\tfrac{1}{2}$ and $\eta=0$) and channel symmetry ($J_{\rm L}=J_{\rm B} \equiv J$, which implies 
$t_{\rm B}=\xi t_{\rm L}$ with $\xi^2\simeq 1+2E_C/U'_d$), whence~\cite{le2004maximized,eq17}
\begin{align}
\nonumber
H_{\rm eff} &= H_0 + J \vec{S}_d \cdot \left ( c_{\alpha\sigma}^{\dagger}\frac{\vec{\boldsymbol{\sigma}}_{\sigma\sigma'}}{2}c_{\alpha\sigma'}\right ) + V_z \hat{T}^z_{\rm B}\left( c_{\alpha\sigma}^{\dagger}\frac{\boldsymbol{\tau}_{\alpha\beta}^z}{2}  c_{\beta\sigma}\right )  \\ 
&+ Q_{\bot}\vec{S}_d \cdot \left ( 
c_{\alpha\sigma}^{\dagger}\vec{\boldsymbol{\sigma}}_{\sigma\sigma'}(\boldsymbol{\tau}_{\alpha\beta}^+\hat{T}^-_{\rm B} + \boldsymbol{\tau}_{\alpha\beta}^-\hat{T}^+_{\rm B})c_{\beta\sigma'} \right) 
%
\label{eq:Heff}
\end{align}
where the Pauli matrices $\boldsymbol{\sigma}^a$  ($\boldsymbol{\tau}^b$) act in spin (channel) space, and a sum over repeated indices is now implied. 

Although initially the coupling constants in Eq.~\ref{eq:Heff} take different values, perturbative scaling~\cite{eq20} shows that the model develops an emergent symmetry $J=V_z=Q_{\bot}$ at an isotropic low-temperature fixed point. Then the RG equations reduce to $dJ/dl=3J^2$, and we have a Kondo scale $T_{\rm K}^{\rm SO(5)}\sim D \exp(-1/3\nu J)$. 

The fixed point has an unusual SO(5) symmetry, which can be seen by writing Eq.~\ref{eq:Heff} in the symmetric form,
\begin{align}
\label{eq:Hso5}
H_{\rm SO(5)} = H_0+ J \sum_{A=1}^{10} J^A M^A \;,
\end{align}
where $J^A=c_{\alpha\sigma}^{\dagger} T_{\alpha\beta\sigma\sigma'}^A c_{\beta\sigma'}$ and $M^A=f_{\alpha\sigma}^{\dagger} T_{\alpha\beta\sigma\sigma'}^A f_{\beta\sigma'}$ in terms of a fermionic  `impurity' operator carrying both `flavor' and spin labels subject to the constraint $f^{\dagger}_{\alpha\sigma}f_{\alpha\sigma}=1$ such that $\hat{S}^a_d=\tfrac{1}{2}f_{\alpha\sigma}^{\dagger}\boldsymbol{\sigma}_{\sigma\sigma'}^a f_{\alpha\sigma'}$ and $\hat{T}^b_{\rm B}=\tfrac{1}{2}f_{\alpha\sigma}^{\dagger}\boldsymbol{\tau}_{\alpha\beta}^b f_{\beta\sigma}$. Here, $\{T^A \}$ are the ten 
generators of SO(5)~\cite{georgi2018lie}, $T^{ab}=-T^{ba}$ (with $a,b=1...5$) satisfying the algebra $[T^{ab},T^{cd}]=-i (\delta_{bc} T^{ad}-\delta_{ac} T^{bd}-\delta_{bd} T^{ac}+\delta_{ad} T^{bc})$ in the 4-dimensional spinor representation,
\begin{align*}
&\tfrac{1}{2}\sigma^{a=1,2,3}\tau^1=T^{a4} \;, \quad
&\tfrac{1}{2}\sigma^1\tau^0=T^{23} \;, \quad
&\tfrac{1}{2}\sigma^3\tau^0=T^{12} \;, \\
&\tfrac{1}{2}\sigma^{a=1,2,3}\tau^2=T^{a5} \;, \quad
&\tfrac{1}{2}\sigma^2\tau^0=T^{31} \;, \quad
&\tfrac{1}{2}\sigma^0\tau^3=T^{45} \;,
\end{align*}
establishing the equivalence between Eqs.~\ref{eq:Heff} and \ref{eq:Hso5}.

We applied the machinery of CFT \cite{affleck1990current,affleck1991critical,affleck1995conformal} to analyze the fixed point properties using the symmetry decomposition U(1)$_{c} \times Z_2 \times$ SO(5)$_1$. Here, U(1) corresponds to the charge sector and $Z_2$ is an Ising model. 
The primary fields of the SO(5)$_1$ theory consist of a singlet with scaling dimension 0, a spinor with scaling dimension $\frac{5}{16}$, and a vector with scaling dimension $\frac{1}{2}$~\cite{francesco2012conformal}. The SO(5) fixed point can be obtained by fusion with the spinor under which the impurity transforms. 

\begin{figure}[t!]
\includegraphics[width=8.5cm]{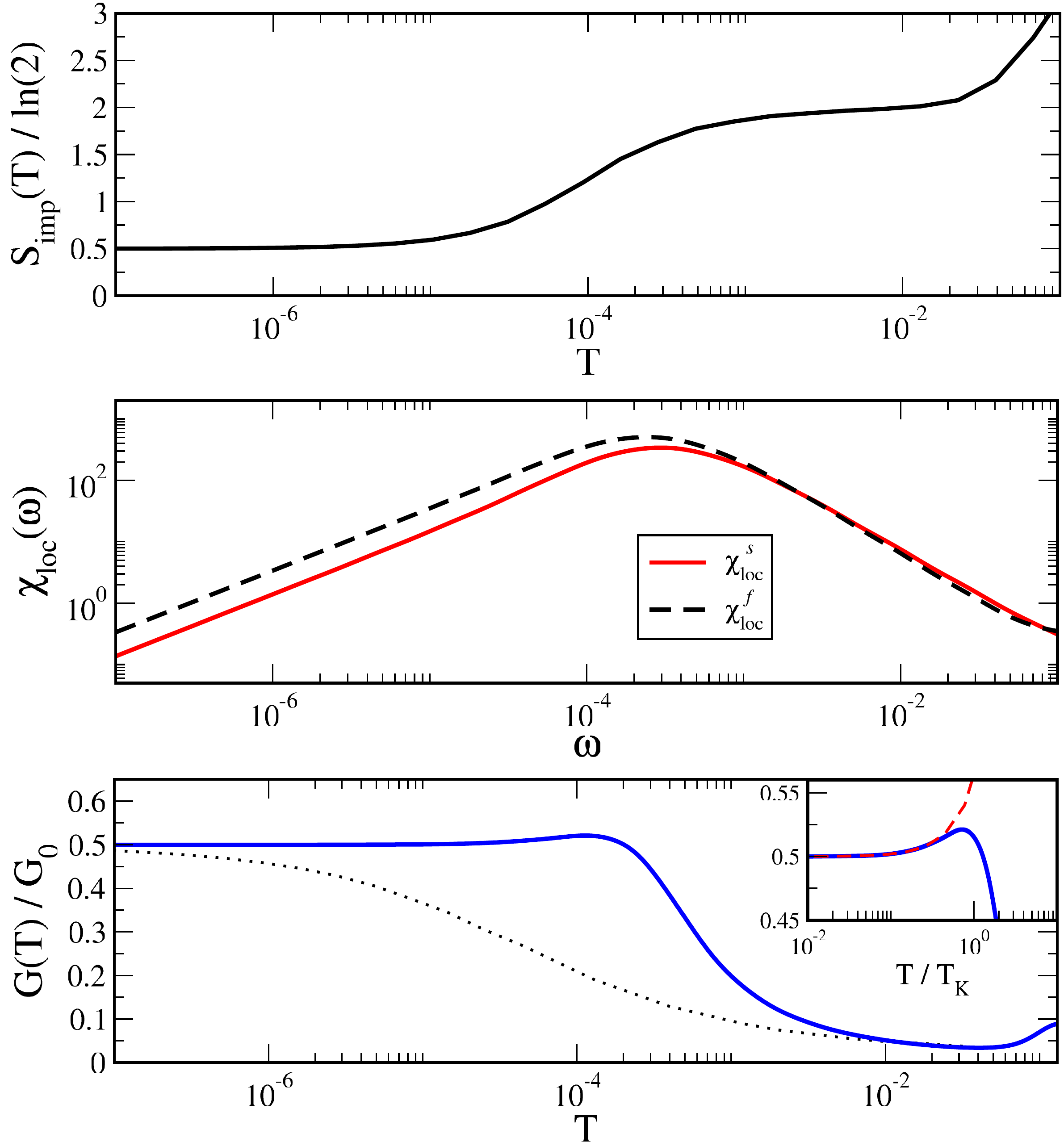}
  \caption{Physical properties of the SO(5) Kondo effect, obtained by NRG for $n_{\rm B}=\tfrac{1}{2}$, $\eta=0$, $U_d=0.3$, $E_C=0.1$, $t_{\rm L}=0.085$, and $t_{\rm B}\simeq \xi t_{\rm L}=0.1$. (a) Impurity contribution to entropy $S_{\rm imp}(T)$, showing 
  partial quenching of the entangled spin and flavor degrees of freedom on the scale of the Kondo temperature $T_{\rm K}\sim 10^{-4}$. For $T_{\rm K}\ll T \ll E_C$, free impurity spin and flavor give a $\ln(4)$ entropy, while $S_{\rm imp}(T)=\tfrac{1}{2}\ln(2)$ for $T\ll T_{\rm K}$, characteristic of the free Majorana fermion at the SO(5) fixed point. (b) $T=0$ local spin and flavor dynamical susceptibilities, both showing apparent FL-like behavior $\chi_{\rm loc}(\omega)\sim \omega$ for $\omega\ll T_{\rm K}$. (c) Linear response conductance through the dot $G(T)/G_0$ (blue line), with $G=\tfrac{1}{2}G_0$ at $T=0$, and leading behavior $G(T)-G(0)\sim +(T/T_{\rm K})^{3/2}G_0$ (inset, dashed line). The standard spin-2CK conductance lineshape is given for comparison as the dotted line.}
  \label{fig:propertiesSO5}
\end{figure}

The finite size spectrum provides a means of characterizing the fixed point. For an effective 1D system of length $L$, the energies ($E$) in units of $2\pi v_F/L$, and corresponding degeneracies ($\#$), can be determined from CFT. 
We find~\cite{SM} $(E, \#) = (0,2) ; (\frac{1}{8},4) ; (\frac{1}{2},10) ; (\frac{5}{8},12) ; (1,26) ; ...$, consistent with our NRG results, and establishing the new SO(5) fixed point as NFL.  Interestingly, this spectrum is identical to that of the standard s-2CK model \cite{affleck1991critical}. 
The entropy at the fixed point is given in terms of the modular S-matrix within CFT \cite{affleck1990current,affleck1991critical,affleck1995conformal}, and here yields $S_{\rm imp}=\tfrac{1}{2}\ln(2)$, consistent with NRG (top panel, Fig.~\ref{fig:propertiesSO5}); again reminiscent of s-2CK.

However, differences from the standard s-2CK picture can be seen in dynamical quantities such as the local susceptibilities and conductance -- see middle and bottom panels in Fig.~\ref{fig:propertiesSO5}. Since the impurity spin $\vec{S}$ and pseudospin $\vec{T}$ operators are absorbed into the conduction electrons at the strong coupling fixed point, they must transform among the 10 generators of SO(5). But such fields occur only as \textit{descendants} in SO(5)$_1$, and so spin-spin correlation functions appear FL-like, $\chi_{\rm loc}^s(\omega)\sim \omega$ (similarly for flavor susceptibility). This contrasts to the regular $k$-channel Kondo effect: the spin SU(2)$_k$ theory contains a vector field which transforms as the 3 components of the impurity spin, with scaling dimension $\tfrac{2}{2+k}$, which leads to anomalous NFL properties in the spin susceptibility $\chi_{\rm loc}^s\sim \omega^{-\frac{k-2}{k+2}}$. 
The SO(5) point appears to have been missed in Ref.~\cite{le2004maximized} because of the apparent FL scaling of its susceptibilities. However, the $\tfrac{1}{2}\ln(2)$ residual entropy is a clear NFL signature. Furthermore, we find a non-monotonic conductance, with a NFL leading power law $G(T)-G(0) \sim +T^{3/2}$. This contrasts to s-2CK conductance which approaches its fixed point value as $-\sqrt{T}$ \cite{oreg2003two,pustilnik2004quantum,sela2011exact,mitchell2012universal}, or $-T^2$ FL conductance for 1CK \cite{pustilnik2004kondo}. 


To gain further insight, we expand on the bosonization and refermionization techniques \cite{von1998bosonization} developed by Emery and Kivelson (EK) for the s-2CK model \cite{emery1992mapping}, and include the coupling to the flavour degree of freedom. This method allows us to express a spin and flavor anisotropic version of Eq.~\ref{eq:Heff} in terms of local fermions $d\propto S_{{\rm{d}}}^-$ and $a\propto T_{{\rm{B}}}^-$, relating to impurity spin and flavor degrees of freedom, with  corresponding Majorana operators $d_{+}=\tfrac{1}{\sqrt{2}}(d^{\dagger}+ d)$ and $d_{-}=\tfrac{1}{\sqrt{2}i}(d^{\dagger}- d)$, and similarly for $a$, as well as a 1D bulk fermionic `spin-flavor' field for the conduction electrons denoted $\psi_{sf}(x)$, with Majorana components $\chi_{+}=\tfrac{1}{\sqrt{2}}(\psi_{\rm{sf}}^{\dagger}(0)+\psi_{\rm{sf}}(0))$, $\chi_{-}=\frac{1}{\sqrt{2}i}(\psi_{\rm{sf}}^{\dagger}(0)-\psi_{\rm{sf}}(0))$. The resulting EK Hamiltonian takes the simple quadratic form,
\begin{equation}
\label{eq:EK}
H_{\rm EK} = H_{0}+iJ_{\perp}d_{-}\chi_{+} -2iQ_{\perp}d_{+}a_{-}. 
\end{equation}
Details of the derivation are given in the Supplemental Material \cite{SM}. The $J_{\bot}$ term is the usual EK form of the s-2CK interaction. The spin-flavour coupling term $Q_{\perp}$  couples and gaps out the pair $d_{+}$ and $a_{-}$. Unlike in the s-2CK model where $d_+$ remains decoupled, here we see that it is $a_+$ that is free at the SO(5) fixed point, and is responsible for the $\tfrac{1}{2}\ln(2)$ residual entropy. The fixed point properties of Eq.~\ref{eq:EK} describe the physical quantum dot system because the artificial spin/flavor anisotropies used to obtain it are RG irrelevant.


\textit{Stability of SO(5) Kondo.--} We consider the effect~of symmetry-breaking perturbations at the SO(5) point.

Channel asymmetry, corresponding to $t_{\rm B}\ne \xi t_{\rm L}$ in the bare model, generates an extra term in Eq.~\ref{eq:Heff} given by  
 $\delta H_{\rm{ ch}}=J_{-}\vec{S}_d\cdot (c_{\alpha\sigma}^{\dagger} \vec{\boldsymbol{\sigma}}_{\sigma\sigma'}\boldsymbol{\tau}_{\alpha\beta}^z c_{\beta\sigma'})$ with $J_{-}\propto J_B-J_L$. 
Under the EK mapping, this becomes $\delta H_{\rm{ ch}}=-iJ_{-} d_{+}\chi_{-}$
 as in the s-2CK model. But in contrast to the s-2CK model, the SO(5) point is not destabilized by this perturbation because the $d_+$ Majorana involved in $J_{-}$ is already gapped out by the spin-flavor coupling $Q_{\bot}$ in Eq.~\ref{eq:EK}. The $a_+$ Majorana remains free. 
 Breaking spin symmetry by applying a dot magnetic field $\delta H_s=B\hat{S}_d^z =-iBd_+ d_-$ is similarly irrelevant at the SO(5) fixed point. The NFL physics is therefore robust to breaking channel and spin symmetries. This is directly confirmed by NRG~\cite{SM}.
 
 Ph symmetry is broken by $\eta \ne 0$ for $n_{\rm B}=\tfrac{1}{2}$. Performing the SWT yields an additional contribution to Eq.~\ref{eq:Heff} of the form~\cite{eq17} $\delta H_{\rm ph} =  \tfrac{1}{2}V_{\bot}\sum_{b=1,2} \hat{T}_{\rm B}^b (c_{\alpha\sigma}^{\dagger}\boldsymbol{\tau}_{\alpha\beta}^b c_{\beta\sigma})  + Q_z\sum_{a=1,2,3} \hat{S}_d^a \hat{T}_{\rm B}^3 (c_{\alpha\sigma}^{\dagger} \boldsymbol{\sigma}_{\sigma\sigma'}^a \boldsymbol{\tau}_{\alpha\beta}^3 c_{\beta\sigma'})$ where $V_\perp , Q_z \propto \eta$. This perturbation contains an additional 5 generators, which together with the 10 from SO(5) form the defining representation of SU(4). Indeed, under RG the system flows to a fully isotropic SU(4) FL fixed point, as discussed in Refs.~\cite{borda20034,le2004maximized,le2003smearing}, with zero residual entropy. Breaking ph symmetry therefore destabilizes the NFL SO(5) fixed point, with an emergent FL crossover scale $T^*\sim \eta^2$~\cite{SM}.
Unusually then, lowering the symmetry of the bare model by introducing finite $\eta$ leads to a low-energy SU(4) fixed point with higher symmetry than the SO(5) fixed point obtained at $\eta=0$. 
Applying the EK mapping, we obtain~\cite{SM}  $\delta H_{\rm ph} =  -iV_{\perp}a_{+}\chi_{-}$.  This is an RG relevant term with scaling dimension $\tfrac{1}{2}$: the previously free $a_{+}$ Majorana is now coupled to the $\chi_{-}$ field, quenching the $\tfrac{1}{2}\ln(2)$ entropy and leading to an FL state, with $\chi_{\rm loc}^{s,f}\sim \omega$ and $G(T)-G(0)\sim T^2$~\cite{inprep}.


\begin{figure}[h!]
\includegraphics[width=8.5cm]{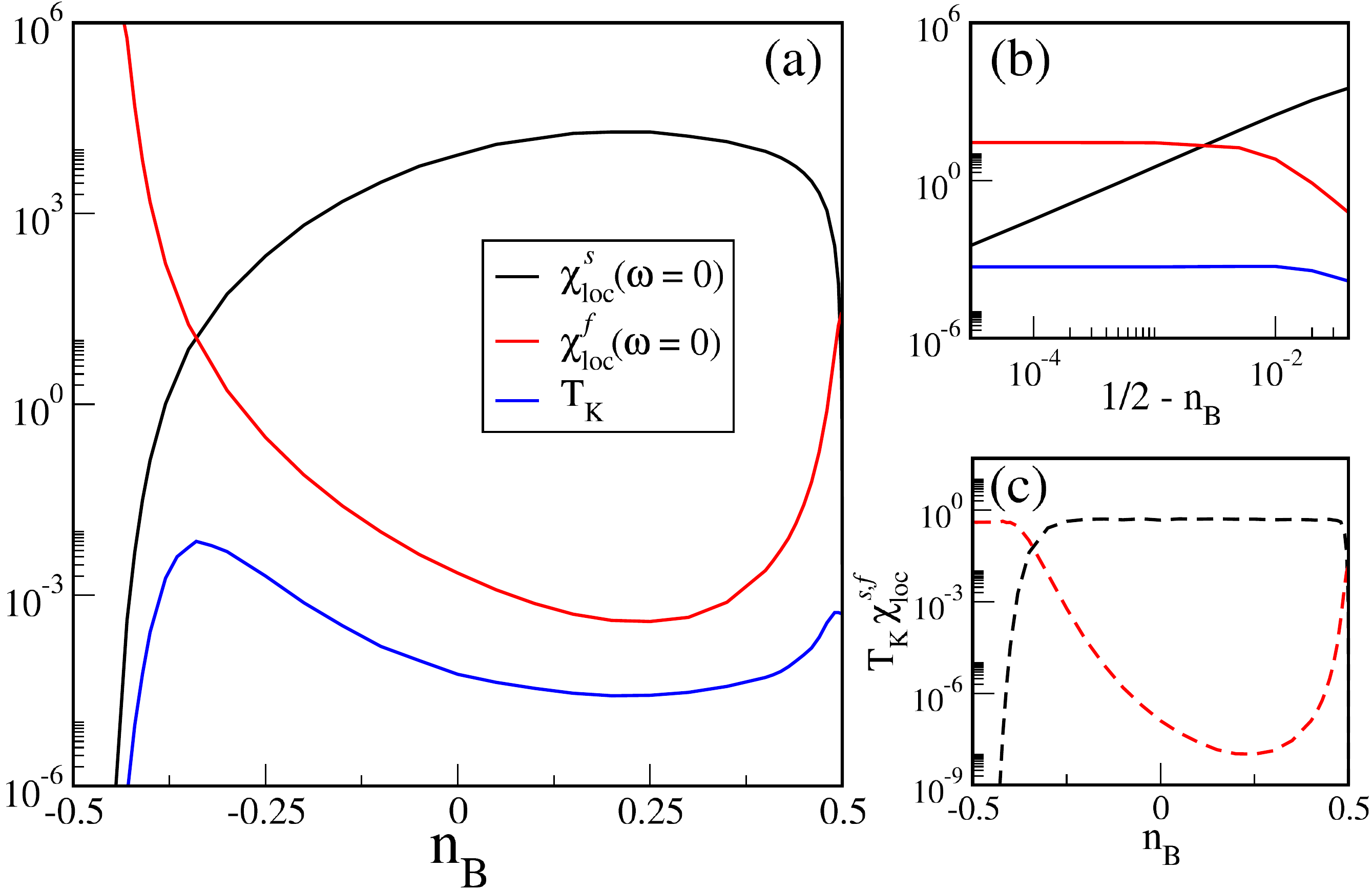}
  \caption{Evolution along the black NFL line in Fig.~\ref{fig:phase_diagram}. (a) $T=\omega=0$ susceptibilities $\chi_{\rm loc}^s$ (black), $\chi_{\rm loc}^f$ (red), and Kondo temperature $T_{\rm K}$ (blue). Flavor fluctuations are suppressed in the box Coulomb blockade regime, while spin fluctuations are suppressed as the dot local moment is lost for $n_{\rm B} \to -\tfrac{1}{2}$. (b) Behavior near $n_{\rm B}=\tfrac{1}{2}$ showing $\chi_{\rm loc}^s(0) \sim (\tfrac{1}{2}-n_{\rm B})^2$ (consistent with $\chi_{\rm loc}^s(\omega) \sim \omega$),  while $\chi_{\rm loc}^f$ remains finite. (c) $T_{\rm K}\chi_{\rm loc}^s$ (black dashed) and $T_{\rm K}\chi_{\rm loc}^f$ (red dashed) showing crossover from spin-flavor to spin to flavor 2CK as $n_{\rm B}$ is varied from $+\tfrac{1}{2}$ to $-\tfrac{1}{2}$.
  }
  \label{fig:susc}
\end{figure}

\textit{Full phase diagram.--} We now explore the entire $(n_{\rm B}$,$\eta)$ plane using NRG, focusing on the quantum critical lines along which NFL physics can be realized in experiment -- see Fig.~\ref{fig:phase_diagram}. In practice, we tune $\eta$ for a given $n_{\rm B}$ to find the critical point, which we identify in NRG from its characteristic $\tfrac{1}{2}\ln(2)$ residual entropy. 

An effective flavor field is generated on moving away from $n_{\rm B} = \tfrac{1}{2}$. The resulting perturbation $\delta H_{\rm{f}} = B_{\rm{f}} \hat{T}_{\rm B}^z$, with $B_{\rm{f}}=E_C(1-2n_{\rm B})$, breaks ph symmetry and is therefore RG relevant at the SO(5) fixed point. However, the two sources of ph asymmetry from $\eta\ne 0$ and $n_{\rm B}\ne \tfrac{1}{2}$ can cancel out (close to the SO(5) point this arises along the line $2Q_{\perp}V_{\perp}=J_{-}B_{{\rm{f}}}$ \cite{inprep}). NRG results confirm the continuous evolution of the NFL state on moving from $n_{\rm B}=\tfrac{1}{2}$ into the box Coulomb blockade regime centered on $n_{\rm B}=0$. In particular, the spin and flavor susceptibilities along the NFL lines in Fig.~\ref{fig:susc} show the crossover from spin-flavor Kondo to s-2CK.

The NFL line with $t_{\rm B}/t_{\rm L}= \zeta$ (blue line, Fig.~\ref{fig:phase_diagram}) is invariant to the ph transformation $n_{\rm B}\to -n_{\rm B}$ and $\eta\to -\eta$, and smoothly connects spin-flavor Kondo at all half-odd-integer $n_{\rm B}$ with s-2CK at all integer $n_{\rm B}$. 
For $t_{\rm B}/t_{\rm L} < \zeta$ (green and pink lines, Fig.~\ref{fig:phase_diagram}), the NFL lines terminate, signalling that the effective ph asymmetry from $n_{\rm B}$ can no longer be compensated by tuning $\eta$.

By contrast, for $t_{\rm B}/t_{\rm L} > \zeta$ (red and black lines, Fig.~\ref{fig:phase_diagram}), the NFL lines continue into an unexpected region of the phase diagram where $|\eta|/U_d >\tfrac{1}{2}$. Here spin fluctuations are suppressed since the dot no longer hosts a local moment, but flavor fluctuations are enhanced (see Fig.~\ref{fig:susc}). The NFL lines in this regime diverge with $\eta \to \pm \infty$ as $n_{\rm B}\to \pm\tfrac{1}{2}$, and f-2CK dominates. Along the crossover from s-2CK to f-2CK, spin and flavor fluctuations become equally balanced at the 
dot charge degeneracy point $|\eta|=\tfrac{1}{2}U_d$. The NFL state at this point develops at a strongly enhanced Kondo temperature (Fig.~\ref{fig:susc}), making it particularly well suited to experimental investigation.

\textit{Conclusions.--} 
We revisit a classic model describing quantum dot/box experiments used to probe NFL physics, uncovering a rich range of new physics, including a novel spin-flavor SO(5) Kondo effect. We study the evolution of the NFL line as a function of dot and box gate voltages using a combination of analytical and numerical techniques, showing that the well-known s-2CK effect can continuously transform into the f-2CK or SO(5) Kondo effects. Distinctive experimental signatures of this new physics should be observable in conductance~\cite{inprep}.\\


\begin{acknowledgments}
\emph{Acknowledgments.--} 
We thank Josh Folk for stimulating discussions. AKM and ES thank the Stewart Blusson Quantum Matter Institute (UBC) for travel support. AKM acknowledges funding from the Irish Research Council Laureate Awards 2017/2018 through grant IRCLA/2017/169.  ES acknowledges support from ARO (W911NF-20-1-0013), the Israel Science Foundation grant number 154/19 and US-Israel Binational Science Foundation (Grant No. 2016255). IA acknowledges support from NSERC Discovery Grant 04033-2016. 
\end{acknowledgments}


%



\begin{thebibliography}{62}%
\makeatletter
\providecommand \@ifxundefined [1]{%
 \@ifx{#1\undefined}
}%
\providecommand \@ifnum [1]{%
 \ifnum #1\expandafter \@firstoftwo
 \else \expandafter \@secondoftwo
 \fi
}%
\providecommand \@ifx [1]{%
 \ifx #1\expandafter \@firstoftwo
 \else \expandafter \@secondoftwo
 \fi
}%
\providecommand \natexlab [1]{#1}%
\providecommand \enquote  [1]{``#1''}%
\providecommand \bibnamefont  [1]{#1}%
\providecommand \bibfnamefont [1]{#1}%
\providecommand \citenamefont [1]{#1}%
\providecommand \href@noop [0]{\@secondoftwo}%
\providecommand \href [0]{\begingroup \@sanitize@url \@href}%
\providecommand \@href[1]{\@@startlink{#1}\@@href}%
\providecommand \@@href[1]{\endgroup#1\@@endlink}%
\providecommand \@sanitize@url [0]{\catcode `\\12\catcode `\$12\catcode
  `\&12\catcode `\#12\catcode `\^12\catcode `\_12\catcode `\%12\relax}%
\providecommand \@@startlink[1]{}%
\providecommand \@@endlink[0]{}%
\providecommand \url  [0]{\begingroup\@sanitize@url \@url }%
\providecommand \@url [1]{\endgroup\@href {#1}{\urlprefix }}%
\providecommand \urlprefix  [0]{URL }%
\providecommand \Eprint [0]{\href }%
\providecommand \doibase [0]{http://dx.doi.org/}%
\providecommand \selectlanguage [0]{\@gobble}%
\providecommand \bibinfo  [0]{\@secondoftwo}%
\providecommand \bibfield  [0]{\@secondoftwo}%
\providecommand \translation [1]{[#1]}%
\providecommand \BibitemOpen [0]{}%
\providecommand \bibitemStop [0]{}%
\providecommand \bibitemNoStop [0]{.\EOS\space}%
\providecommand \EOS [0]{\spacefactor3000\relax}%
\providecommand \BibitemShut  [1]{\csname bibitem#1\endcsname}%
\let\auto@bib@innerbib\@empty
\bibitem [{\citenamefont {Sohn}\ \emph {et~al.}(2013)\citenamefont {Sohn},
  \citenamefont {Kouwenhoven},\ and\ \citenamefont
  {Sch{\"o}n}}]{sohn2013mesoscopic}%
  \BibitemOpen
  \bibfield  {author} {\bibinfo {author} {\bibfnamefont {L.~L.}\ \bibnamefont
  {Sohn}}, \bibinfo {author} {\bibfnamefont {L.~P.}\ \bibnamefont
  {Kouwenhoven}}, \ and\ \bibinfo {author} {\bibfnamefont {G.}~\bibnamefont
  {Sch{\"o}n}},\ }\href@noop {} {\emph {\bibinfo {title} {Mesoscopic electron
  transport}}},\ Vol.\ \bibinfo {volume} {345}\ (\bibinfo  {publisher}
  {Springer Science \& Business Media},\ \bibinfo {year} {2013})\BibitemShut
  {NoStop}%
\bibitem [{\citenamefont {Pustilnik}\ and\ \citenamefont
  {Glazman}(2004)}]{pustilnik2004kondo}%
  \BibitemOpen
  \bibfield  {author} {\bibinfo {author} {\bibfnamefont {M.}~\bibnamefont
  {Pustilnik}}\ and\ \bibinfo {author} {\bibfnamefont {L.}~\bibnamefont
  {Glazman}},\ }\href@noop {} {\bibfield  {journal} {\bibinfo  {journal}
  {Journal of Physics: Condensed Matter}\ }\textbf {\bibinfo {volume} {16}},\
  \bibinfo {pages} {R513} (\bibinfo {year} {2004})}\BibitemShut {NoStop}%
\bibitem [{\citenamefont {Goldhaber-Gordon}\ \emph {et~al.}(1998)\citenamefont
  {Goldhaber-Gordon}, \citenamefont {Shtrikman}, \citenamefont {Mahalu},
  \citenamefont {Abusch-Magder}, \citenamefont {Meirav},\ and\ \citenamefont
  {Kastner}}]{goldhaber1998kondo}%
  \BibitemOpen
  \bibfield  {author} {\bibinfo {author} {\bibfnamefont {D.}~\bibnamefont
  {Goldhaber-Gordon}}, \bibinfo {author} {\bibfnamefont {H.}~\bibnamefont
  {Shtrikman}}, \bibinfo {author} {\bibfnamefont {D.}~\bibnamefont {Mahalu}},
  \bibinfo {author} {\bibfnamefont {D.}~\bibnamefont {Abusch-Magder}}, \bibinfo
  {author} {\bibfnamefont {U.}~\bibnamefont {Meirav}}, \ and\ \bibinfo {author}
  {\bibfnamefont {M.}~\bibnamefont {Kastner}},\ }\href@noop {} {\bibfield
  {journal} {\bibinfo  {journal} {Nature}\ }\textbf {\bibinfo {volume} {391}},\
  \bibinfo {pages} {156} (\bibinfo {year} {1998})}\BibitemShut {NoStop}%
\bibitem [{\citenamefont {Cronenwett}\ \emph {et~al.}(1998)\citenamefont
  {Cronenwett}, \citenamefont {Oosterkamp},\ and\ \citenamefont
  {Kouwenhoven}}]{cronenwett1998tunable}%
  \BibitemOpen
  \bibfield  {author} {\bibinfo {author} {\bibfnamefont {S.~M.}\ \bibnamefont
  {Cronenwett}}, \bibinfo {author} {\bibfnamefont {T.~H.}\ \bibnamefont
  {Oosterkamp}}, \ and\ \bibinfo {author} {\bibfnamefont {L.~P.}\ \bibnamefont
  {Kouwenhoven}},\ }\href@noop {} {\bibfield  {journal} {\bibinfo  {journal}
  {Science}\ }\textbf {\bibinfo {volume} {281}},\ \bibinfo {pages} {540}
  (\bibinfo {year} {1998})}\BibitemShut {NoStop}%
\bibitem [{\citenamefont {Van~der Wiel}\ \emph {et~al.}(2000)\citenamefont
  {Van~der Wiel}, \citenamefont {De~Franceschi}, \citenamefont {Fujisawa},
  \citenamefont {Elzerman}, \citenamefont {Tarucha},\ and\ \citenamefont
  {Kouwenhoven}}]{van2000kondo}%
  \BibitemOpen
  \bibfield  {author} {\bibinfo {author} {\bibfnamefont {W.}~\bibnamefont
  {Van~der Wiel}}, \bibinfo {author} {\bibfnamefont {S.}~\bibnamefont
  {De~Franceschi}}, \bibinfo {author} {\bibfnamefont {T.}~\bibnamefont
  {Fujisawa}}, \bibinfo {author} {\bibfnamefont {J.}~\bibnamefont {Elzerman}},
  \bibinfo {author} {\bibfnamefont {S.}~\bibnamefont {Tarucha}}, \ and\
  \bibinfo {author} {\bibfnamefont {L.}~\bibnamefont {Kouwenhoven}},\
  }\href@noop {} {\bibfield  {journal} {\bibinfo  {journal} {science}\ }\textbf
  {\bibinfo {volume} {289}},\ \bibinfo {pages} {2105} (\bibinfo {year}
  {2000})}\BibitemShut {NoStop}%
\bibitem [{\citenamefont {Hewson}(1997)}]{hewson1997kondo}%
  \BibitemOpen
  \bibfield  {author} {\bibinfo {author} {\bibfnamefont {A.~C.}\ \bibnamefont
  {Hewson}},\ }\href@noop {} {\emph {\bibinfo {title} {{The Kondo Problem to
  Heavy Fermions}}}}\ (\bibinfo  {publisher} {Cambridge University Press},\
  \bibinfo {year} {1997})\BibitemShut {NoStop}%
\bibitem [{\citenamefont {Jeong}\ \emph {et~al.}(2001)\citenamefont {Jeong},
  \citenamefont {Chang},\ and\ \citenamefont {Melloch}}]{jeong2001kondo}%
  \BibitemOpen
  \bibfield  {author} {\bibinfo {author} {\bibfnamefont {H.}~\bibnamefont
  {Jeong}}, \bibinfo {author} {\bibfnamefont {A.~M.}\ \bibnamefont {Chang}}, \
  and\ \bibinfo {author} {\bibfnamefont {M.~R.}\ \bibnamefont {Melloch}},\
  }\href@noop {} {\bibfield  {journal} {\bibinfo  {journal} {Science}\ }\textbf
  {\bibinfo {volume} {293}},\ \bibinfo {pages} {2221} (\bibinfo {year}
  {2001})}\BibitemShut {NoStop}%
\bibitem [{\citenamefont {Malecki}\ \emph {et~al.}(2010)\citenamefont
  {Malecki}, \citenamefont {Sela},\ and\ \citenamefont
  {Affleck}}]{malecki2010prospect}%
  \BibitemOpen
  \bibfield  {author} {\bibinfo {author} {\bibfnamefont {J.}~\bibnamefont
  {Malecki}}, \bibinfo {author} {\bibfnamefont {E.}~\bibnamefont {Sela}}, \
  and\ \bibinfo {author} {\bibfnamefont {I.}~\bibnamefont {Affleck}},\
  }\href@noop {} {\bibfield  {journal} {\bibinfo  {journal} {Physical Review
  B}\ }\textbf {\bibinfo {volume} {82}},\ \bibinfo {pages} {205327} (\bibinfo
  {year} {2010})}\BibitemShut {NoStop}%
\bibitem [{\citenamefont {Liang}\ \emph {et~al.}(2002)\citenamefont {Liang},
  \citenamefont {Shores}, \citenamefont {Bockrath}, \citenamefont {Long},\ and\
  \citenamefont {Park}}]{liang2002kondo}%
  \BibitemOpen
  \bibfield  {author} {\bibinfo {author} {\bibfnamefont {W.}~\bibnamefont
  {Liang}}, \bibinfo {author} {\bibfnamefont {M.~P.}\ \bibnamefont {Shores}},
  \bibinfo {author} {\bibfnamefont {M.}~\bibnamefont {Bockrath}}, \bibinfo
  {author} {\bibfnamefont {J.~R.}\ \bibnamefont {Long}}, \ and\ \bibinfo
  {author} {\bibfnamefont {H.}~\bibnamefont {Park}},\ }\href@noop {} {\bibfield
   {journal} {\bibinfo  {journal} {Nature}\ }\textbf {\bibinfo {volume}
  {417}},\ \bibinfo {pages} {725} (\bibinfo {year} {2002})}\BibitemShut
  {NoStop}%
\bibitem [{\citenamefont {Mitchell}\ \emph {et~al.}(2017)\citenamefont
  {Mitchell}, \citenamefont {Pedersen}, \citenamefont {Hedeg{\aa}rd},\ and\
  \citenamefont {Paaske}}]{mitchell2017kondo}%
  \BibitemOpen
  \bibfield  {author} {\bibinfo {author} {\bibfnamefont {A.~K.}\ \bibnamefont
  {Mitchell}}, \bibinfo {author} {\bibfnamefont {K.~G.}\ \bibnamefont
  {Pedersen}}, \bibinfo {author} {\bibfnamefont {P.}~\bibnamefont
  {Hedeg{\aa}rd}}, \ and\ \bibinfo {author} {\bibfnamefont {J.}~\bibnamefont
  {Paaske}},\ }\href@noop {} {\bibfield  {journal} {\bibinfo  {journal} {Nature
  communications}\ }\textbf {\bibinfo {volume} {8}},\ \bibinfo {pages} {1}
  (\bibinfo {year} {2017})}\BibitemShut {NoStop}%
\bibitem [{\citenamefont {L{\'o}pez}\ \emph {et~al.}(2005)\citenamefont
  {L{\'o}pez}, \citenamefont {S{\'a}nchez}, \citenamefont {Lee}, \citenamefont
  {Choi}, \citenamefont {Simon},\ and\ \citenamefont
  {Le~Hur}}]{lopez2005probing}%
  \BibitemOpen
  \bibfield  {author} {\bibinfo {author} {\bibfnamefont {R.}~\bibnamefont
  {L{\'o}pez}}, \bibinfo {author} {\bibfnamefont {D.}~\bibnamefont
  {S{\'a}nchez}}, \bibinfo {author} {\bibfnamefont {M.}~\bibnamefont {Lee}},
  \bibinfo {author} {\bibfnamefont {M.-S.}\ \bibnamefont {Choi}}, \bibinfo
  {author} {\bibfnamefont {P.}~\bibnamefont {Simon}}, \ and\ \bibinfo {author}
  {\bibfnamefont {K.}~\bibnamefont {Le~Hur}},\ }\href@noop {} {\bibfield
  {journal} {\bibinfo  {journal} {Physical Review B}\ }\textbf {\bibinfo
  {volume} {71}},\ \bibinfo {pages} {115312} (\bibinfo {year}
  {2005})}\BibitemShut {NoStop}%
\bibitem [{\citenamefont {Sasaki}\ \emph {et~al.}(2000)\citenamefont {Sasaki},
  \citenamefont {De~Franceschi}, \citenamefont {Elzerman}, \citenamefont
  {Van~der Wiel}, \citenamefont {Eto}, \citenamefont {Tarucha},\ and\
  \citenamefont {Kouwenhoven}}]{sasaki2000kondo}%
  \BibitemOpen
  \bibfield  {author} {\bibinfo {author} {\bibfnamefont {S.}~\bibnamefont
  {Sasaki}}, \bibinfo {author} {\bibfnamefont {S.}~\bibnamefont
  {De~Franceschi}}, \bibinfo {author} {\bibfnamefont {J.}~\bibnamefont
  {Elzerman}}, \bibinfo {author} {\bibfnamefont {W.}~\bibnamefont {Van~der
  Wiel}}, \bibinfo {author} {\bibfnamefont {M.}~\bibnamefont {Eto}}, \bibinfo
  {author} {\bibfnamefont {S.}~\bibnamefont {Tarucha}}, \ and\ \bibinfo
  {author} {\bibfnamefont {L.}~\bibnamefont {Kouwenhoven}},\ }\href@noop {}
  {\bibfield  {journal} {\bibinfo  {journal} {Nature}\ }\textbf {\bibinfo
  {volume} {405}},\ \bibinfo {pages} {764} (\bibinfo {year}
  {2000})}\BibitemShut {NoStop}%
\bibitem [{\citenamefont {Paaske}\ \emph {et~al.}(2006)\citenamefont {Paaske},
  \citenamefont {Rosch}, \citenamefont {W{\"o}lfle}, \citenamefont {Mason},
  \citenamefont {Marcus},\ and\ \citenamefont {Nyg{\aa}rd}}]{paaske2006non}%
  \BibitemOpen
  \bibfield  {author} {\bibinfo {author} {\bibfnamefont {J.}~\bibnamefont
  {Paaske}}, \bibinfo {author} {\bibfnamefont {A.}~\bibnamefont {Rosch}},
  \bibinfo {author} {\bibfnamefont {P.}~\bibnamefont {W{\"o}lfle}}, \bibinfo
  {author} {\bibfnamefont {N.}~\bibnamefont {Mason}}, \bibinfo {author}
  {\bibfnamefont {C.}~\bibnamefont {Marcus}}, \ and\ \bibinfo {author}
  {\bibfnamefont {J.}~\bibnamefont {Nyg{\aa}rd}},\ }\href@noop {} {\bibfield
  {journal} {\bibinfo  {journal} {Nature Physics}\ }\textbf {\bibinfo {volume}
  {2}},\ \bibinfo {pages} {460} (\bibinfo {year} {2006})}\BibitemShut {NoStop}%
\bibitem [{\citenamefont {Mitchell}\ \emph {et~al.}(2009)\citenamefont
  {Mitchell}, \citenamefont {Jarrold},\ and\ \citenamefont
  {Logan}}]{mitchell2009quantum}%
  \BibitemOpen
  \bibfield  {author} {\bibinfo {author} {\bibfnamefont {A.~K.}\ \bibnamefont
  {Mitchell}}, \bibinfo {author} {\bibfnamefont {T.~F.}\ \bibnamefont
  {Jarrold}}, \ and\ \bibinfo {author} {\bibfnamefont {D.~E.}\ \bibnamefont
  {Logan}},\ }\href@noop {} {\bibfield  {journal} {\bibinfo  {journal}
  {Physical Review B}\ }\textbf {\bibinfo {volume} {79}},\ \bibinfo {pages}
  {085124} (\bibinfo {year} {2009})}\BibitemShut {NoStop}%
\bibitem [{\citenamefont {Keller}\ \emph {et~al.}(2014)\citenamefont {Keller},
  \citenamefont {Amasha}, \citenamefont {Weymann}, \citenamefont {Moca},
  \citenamefont {Rau}, \citenamefont {Katine}, \citenamefont {Shtrikman},
  \citenamefont {Zar{\'a}nd},\ and\ \citenamefont
  {Goldhaber-Gordon}}]{keller2014emergent}%
  \BibitemOpen
  \bibfield  {author} {\bibinfo {author} {\bibfnamefont {A.}~\bibnamefont
  {Keller}}, \bibinfo {author} {\bibfnamefont {S.}~\bibnamefont {Amasha}},
  \bibinfo {author} {\bibfnamefont {I.}~\bibnamefont {Weymann}}, \bibinfo
  {author} {\bibfnamefont {C.}~\bibnamefont {Moca}}, \bibinfo {author}
  {\bibfnamefont {I.}~\bibnamefont {Rau}}, \bibinfo {author} {\bibfnamefont
  {J.}~\bibnamefont {Katine}}, \bibinfo {author} {\bibfnamefont
  {H.}~\bibnamefont {Shtrikman}}, \bibinfo {author} {\bibfnamefont
  {G.}~\bibnamefont {Zar{\'a}nd}}, \ and\ \bibinfo {author} {\bibfnamefont
  {D.}~\bibnamefont {Goldhaber-Gordon}},\ }\href@noop {} {\bibfield  {journal}
  {\bibinfo  {journal} {Nature Physics}\ }\textbf {\bibinfo {volume} {10}},\
  \bibinfo {pages} {145} (\bibinfo {year} {2014})}\BibitemShut {NoStop}%
\bibitem [{\citenamefont {Borda}\ \emph {et~al.}(2003)\citenamefont {Borda},
  \citenamefont {Zar{\'a}nd}, \citenamefont {Hofstetter}, \citenamefont
  {Halperin},\ and\ \citenamefont {Von~Delft}}]{borda20034}%
  \BibitemOpen
  \bibfield  {author} {\bibinfo {author} {\bibfnamefont {L.}~\bibnamefont
  {Borda}}, \bibinfo {author} {\bibfnamefont {G.}~\bibnamefont {Zar{\'a}nd}},
  \bibinfo {author} {\bibfnamefont {W.}~\bibnamefont {Hofstetter}}, \bibinfo
  {author} {\bibfnamefont {B.}~\bibnamefont {Halperin}}, \ and\ \bibinfo
  {author} {\bibfnamefont {J.}~\bibnamefont {Von~Delft}},\ }\href@noop {}
  {\bibfield  {journal} {\bibinfo  {journal} {Physical review letters}\
  }\textbf {\bibinfo {volume} {90}},\ \bibinfo {pages} {026602} (\bibinfo
  {year} {2003})}\BibitemShut {NoStop}%
\bibitem [{\citenamefont {Galpin}\ \emph {et~al.}(2005)\citenamefont {Galpin},
  \citenamefont {Logan},\ and\ \citenamefont
  {Krishnamurthy}}]{galpin2005quantum}%
  \BibitemOpen
  \bibfield  {author} {\bibinfo {author} {\bibfnamefont {M.~R.}\ \bibnamefont
  {Galpin}}, \bibinfo {author} {\bibfnamefont {D.~E.}\ \bibnamefont {Logan}}, \
  and\ \bibinfo {author} {\bibfnamefont {H.}~\bibnamefont {Krishnamurthy}},\
  }\href@noop {} {\bibfield  {journal} {\bibinfo  {journal} {Physical review
  letters}\ }\textbf {\bibinfo {volume} {94}},\ \bibinfo {pages} {186406}
  (\bibinfo {year} {2005})}\BibitemShut {NoStop}%
\bibitem [{\citenamefont {Choi}\ \emph {et~al.}(2005)\citenamefont {Choi},
  \citenamefont {L{\'o}pez},\ and\ \citenamefont {Aguado}}]{choi20054}%
  \BibitemOpen
  \bibfield  {author} {\bibinfo {author} {\bibfnamefont {M.-S.}\ \bibnamefont
  {Choi}}, \bibinfo {author} {\bibfnamefont {R.}~\bibnamefont {L{\'o}pez}}, \
  and\ \bibinfo {author} {\bibfnamefont {R.}~\bibnamefont {Aguado}},\
  }\href@noop {} {\bibfield  {journal} {\bibinfo  {journal} {Physical review
  letters}\ }\textbf {\bibinfo {volume} {95}},\ \bibinfo {pages} {067204}
  (\bibinfo {year} {2005})}\BibitemShut {NoStop}%
\bibitem [{\citenamefont {Anders}\ \emph {et~al.}(2008)\citenamefont {Anders},
  \citenamefont {Logan}, \citenamefont {Galpin},\ and\ \citenamefont
  {Finkelstein}}]{anders2008zero}%
  \BibitemOpen
  \bibfield  {author} {\bibinfo {author} {\bibfnamefont {F.~B.}\ \bibnamefont
  {Anders}}, \bibinfo {author} {\bibfnamefont {D.~E.}\ \bibnamefont {Logan}},
  \bibinfo {author} {\bibfnamefont {M.~R.}\ \bibnamefont {Galpin}}, \ and\
  \bibinfo {author} {\bibfnamefont {G.}~\bibnamefont {Finkelstein}},\
  }\href@noop {} {\bibfield  {journal} {\bibinfo  {journal} {Physical review
  letters}\ }\textbf {\bibinfo {volume} {100}},\ \bibinfo {pages} {086809}
  (\bibinfo {year} {2008})}\BibitemShut {NoStop}%
\bibitem [{\citenamefont {Nozieres}\ and\ \citenamefont
  {Blandin}(1980)}]{nozieres1980kondo}%
  \BibitemOpen
  \bibfield  {author} {\bibinfo {author} {\bibfnamefont {P.}~\bibnamefont
  {Nozieres}}\ and\ \bibinfo {author} {\bibfnamefont {A.}~\bibnamefont
  {Blandin}},\ }\href@noop {} {\bibfield  {journal} {\bibinfo  {journal}
  {Journal de Physique}\ }\textbf {\bibinfo {volume} {41}},\ \bibinfo {pages}
  {193} (\bibinfo {year} {1980})}\BibitemShut {NoStop}%
\bibitem [{\citenamefont {Affleck}\ and\ \citenamefont
  {Ludwig}(1991{\natexlab{a}})}]{affleck1991universal}%
  \BibitemOpen
  \bibfield  {author} {\bibinfo {author} {\bibfnamefont {I.}~\bibnamefont
  {Affleck}}\ and\ \bibinfo {author} {\bibfnamefont {A.~W.}\ \bibnamefont
  {Ludwig}},\ }\href@noop {} {\bibfield  {journal} {\bibinfo  {journal}
  {Physical Review Letters}\ }\textbf {\bibinfo {volume} {67}},\ \bibinfo
  {pages} {161} (\bibinfo {year} {1991}{\natexlab{a}})}\BibitemShut {NoStop}%
\bibitem [{\citenamefont {Affleck}\ and\ \citenamefont
  {Ludwig}(1993)}]{affleck1993exact}%
  \BibitemOpen
  \bibfield  {author} {\bibinfo {author} {\bibfnamefont {I.}~\bibnamefont
  {Affleck}}\ and\ \bibinfo {author} {\bibfnamefont {A.~W.}\ \bibnamefont
  {Ludwig}},\ }\href@noop {} {\bibfield  {journal} {\bibinfo  {journal}
  {Physical Review B}\ }\textbf {\bibinfo {volume} {48}},\ \bibinfo {pages}
  {7297} (\bibinfo {year} {1993})}\BibitemShut {NoStop}%
\bibitem [{\citenamefont {Oreg}\ and\ \citenamefont
  {Goldhaber-Gordon}(2003)}]{oreg2003two}%
  \BibitemOpen
  \bibfield  {author} {\bibinfo {author} {\bibfnamefont {Y.}~\bibnamefont
  {Oreg}}\ and\ \bibinfo {author} {\bibfnamefont {D.}~\bibnamefont
  {Goldhaber-Gordon}},\ }\href@noop {} {\bibfield  {journal} {\bibinfo
  {journal} {Physical review letters}\ }\textbf {\bibinfo {volume} {90}},\
  \bibinfo {pages} {136602} (\bibinfo {year} {2003})}\BibitemShut {NoStop}%
\bibitem [{\citenamefont {Furusaki}\ and\ \citenamefont
  {Matveev}(1995)}]{furusaki1995theory}%
  \BibitemOpen
  \bibfield  {author} {\bibinfo {author} {\bibfnamefont {A.}~\bibnamefont
  {Furusaki}}\ and\ \bibinfo {author} {\bibfnamefont {K.}~\bibnamefont
  {Matveev}},\ }\href@noop {} {\bibfield  {journal} {\bibinfo  {journal}
  {Physical Review B}\ }\textbf {\bibinfo {volume} {52}},\ \bibinfo {pages}
  {16676} (\bibinfo {year} {1995})}\BibitemShut {NoStop}%
\bibitem [{\citenamefont {Matveev}(1995)}]{matveev1995coulomb}%
  \BibitemOpen
  \bibfield  {author} {\bibinfo {author} {\bibfnamefont {K.}~\bibnamefont
  {Matveev}},\ }\href@noop {} {\bibfield  {journal} {\bibinfo  {journal}
  {Physical Review B}\ }\textbf {\bibinfo {volume} {51}},\ \bibinfo {pages}
  {1743} (\bibinfo {year} {1995})}\BibitemShut {NoStop}%
\bibitem [{\citenamefont {Potok}\ \emph {et~al.}(2007)\citenamefont {Potok},
  \citenamefont {Rau}, \citenamefont {Shtrikman}, \citenamefont {Oreg},\ and\
  \citenamefont {Goldhaber-Gordon}}]{potok2007observation}%
  \BibitemOpen
  \bibfield  {author} {\bibinfo {author} {\bibfnamefont {R.}~\bibnamefont
  {Potok}}, \bibinfo {author} {\bibfnamefont {I.}~\bibnamefont {Rau}}, \bibinfo
  {author} {\bibfnamefont {H.}~\bibnamefont {Shtrikman}}, \bibinfo {author}
  {\bibfnamefont {Y.}~\bibnamefont {Oreg}}, \ and\ \bibinfo {author}
  {\bibfnamefont {D.}~\bibnamefont {Goldhaber-Gordon}},\ }\href@noop {}
  {\bibfield  {journal} {\bibinfo  {journal} {Nature}\ }\textbf {\bibinfo
  {volume} {446}},\ \bibinfo {pages} {167} (\bibinfo {year}
  {2007})}\BibitemShut {NoStop}%
\bibitem [{\citenamefont {Keller}\ \emph {et~al.}(2015)\citenamefont {Keller},
  \citenamefont {Peeters}, \citenamefont {Moca}, \citenamefont {Weymann},
  \citenamefont {Mahalu}, \citenamefont {Umansky}, \citenamefont {Zar{\'a}nd},\
  and\ \citenamefont {Goldhaber-Gordon}}]{keller2015universal}%
  \BibitemOpen
  \bibfield  {author} {\bibinfo {author} {\bibfnamefont {A.}~\bibnamefont
  {Keller}}, \bibinfo {author} {\bibfnamefont {L.}~\bibnamefont {Peeters}},
  \bibinfo {author} {\bibfnamefont {C.}~\bibnamefont {Moca}}, \bibinfo {author}
  {\bibfnamefont {I.}~\bibnamefont {Weymann}}, \bibinfo {author} {\bibfnamefont
  {D.}~\bibnamefont {Mahalu}}, \bibinfo {author} {\bibfnamefont
  {V.}~\bibnamefont {Umansky}}, \bibinfo {author} {\bibfnamefont
  {G.}~\bibnamefont {Zar{\'a}nd}}, \ and\ \bibinfo {author} {\bibfnamefont
  {D.}~\bibnamefont {Goldhaber-Gordon}},\ }\href@noop {} {\bibfield  {journal}
  {\bibinfo  {journal} {Nature}\ }\textbf {\bibinfo {volume} {526}},\ \bibinfo
  {pages} {237} (\bibinfo {year} {2015})}\BibitemShut {NoStop}%
\bibitem [{\citenamefont {Iftikhar}\ \emph {et~al.}(2015)\citenamefont
  {Iftikhar}, \citenamefont {Jezouin}, \citenamefont {Anthore}, \citenamefont
  {Gennser}, \citenamefont {Parmentier}, \citenamefont {Cavanna},\ and\
  \citenamefont {Pierre}}]{iftikhar2015two}%
  \BibitemOpen
  \bibfield  {author} {\bibinfo {author} {\bibfnamefont {Z.}~\bibnamefont
  {Iftikhar}}, \bibinfo {author} {\bibfnamefont {S.}~\bibnamefont {Jezouin}},
  \bibinfo {author} {\bibfnamefont {A.}~\bibnamefont {Anthore}}, \bibinfo
  {author} {\bibfnamefont {U.}~\bibnamefont {Gennser}}, \bibinfo {author}
  {\bibfnamefont {F.}~\bibnamefont {Parmentier}}, \bibinfo {author}
  {\bibfnamefont {A.}~\bibnamefont {Cavanna}}, \ and\ \bibinfo {author}
  {\bibfnamefont {F.}~\bibnamefont {Pierre}},\ }\href@noop {} {\bibfield
  {journal} {\bibinfo  {journal} {Nature}\ }\textbf {\bibinfo {volume} {526}},\
  \bibinfo {pages} {233} (\bibinfo {year} {2015})}\BibitemShut {NoStop}%
\bibitem [{\citenamefont {Iftikhar}\ \emph {et~al.}(2018)\citenamefont
  {Iftikhar}, \citenamefont {Anthore}, \citenamefont {Mitchell}, \citenamefont
  {Parmentier}, \citenamefont {Gennser}, \citenamefont {Ouerghi}, \citenamefont
  {Cavanna}, \citenamefont {Mora}, \citenamefont {Simon},\ and\ \citenamefont
  {Pierre}}]{iftikhar2018tunable}%
  \BibitemOpen
  \bibfield  {author} {\bibinfo {author} {\bibfnamefont {Z.}~\bibnamefont
  {Iftikhar}}, \bibinfo {author} {\bibfnamefont {A.}~\bibnamefont {Anthore}},
  \bibinfo {author} {\bibfnamefont {A.}~\bibnamefont {Mitchell}}, \bibinfo
  {author} {\bibfnamefont {F.}~\bibnamefont {Parmentier}}, \bibinfo {author}
  {\bibfnamefont {U.}~\bibnamefont {Gennser}}, \bibinfo {author} {\bibfnamefont
  {A.}~\bibnamefont {Ouerghi}}, \bibinfo {author} {\bibfnamefont
  {A.}~\bibnamefont {Cavanna}}, \bibinfo {author} {\bibfnamefont
  {C.}~\bibnamefont {Mora}}, \bibinfo {author} {\bibfnamefont {P.}~\bibnamefont
  {Simon}}, \ and\ \bibinfo {author} {\bibfnamefont {F.}~\bibnamefont
  {Pierre}},\ }\href@noop {} {\bibfield  {journal} {\bibinfo  {journal}
  {Science}\ }\textbf {\bibinfo {volume} {360}},\ \bibinfo {pages} {1315}
  (\bibinfo {year} {2018})}\BibitemShut {NoStop}%
\bibitem [{\citenamefont {Mitchell}\ \emph {et~al.}(2016)\citenamefont
  {Mitchell}, \citenamefont {Landau}, \citenamefont {Fritz},\ and\
  \citenamefont {Sela}}]{mitchell2016universality}%
  \BibitemOpen
  \bibfield  {author} {\bibinfo {author} {\bibfnamefont {A.~K.}\ \bibnamefont
  {Mitchell}}, \bibinfo {author} {\bibfnamefont {L.}~\bibnamefont {Landau}},
  \bibinfo {author} {\bibfnamefont {L.}~\bibnamefont {Fritz}}, \ and\ \bibinfo
  {author} {\bibfnamefont {E.}~\bibnamefont {Sela}},\ }\href@noop {} {\bibfield
   {journal} {\bibinfo  {journal} {Physical review letters}\ }\textbf {\bibinfo
  {volume} {116}},\ \bibinfo {pages} {157202} (\bibinfo {year}
  {2016})}\BibitemShut {NoStop}%
\bibitem [{\citenamefont {Landau}\ \emph {et~al.}(2018)\citenamefont {Landau},
  \citenamefont {Cornfeld},\ and\ \citenamefont {Sela}}]{landau2018charge}%
  \BibitemOpen
  \bibfield  {author} {\bibinfo {author} {\bibfnamefont {L.~A.}\ \bibnamefont
  {Landau}}, \bibinfo {author} {\bibfnamefont {E.}~\bibnamefont {Cornfeld}}, \
  and\ \bibinfo {author} {\bibfnamefont {E.}~\bibnamefont {Sela}},\ }\href@noop
  {} {\bibfield  {journal} {\bibinfo  {journal} {Physical review letters}\
  }\textbf {\bibinfo {volume} {120}},\ \bibinfo {pages} {186801} (\bibinfo
  {year} {2018})}\BibitemShut {NoStop}%
\bibitem [{\citenamefont {Nguyen}\ and\ \citenamefont
  {Kiselev}(2020)}]{nguyen2020thermoelectric}%
  \BibitemOpen
  \bibfield  {author} {\bibinfo {author} {\bibfnamefont {T.}~\bibnamefont
  {Nguyen}}\ and\ \bibinfo {author} {\bibfnamefont {M.}~\bibnamefont
  {Kiselev}},\ }\href@noop {} {\bibfield  {journal} {\bibinfo  {journal}
  {Physical Review Letters}\ }\textbf {\bibinfo {volume} {125}},\ \bibinfo
  {pages} {026801} (\bibinfo {year} {2020})}\BibitemShut {NoStop}%
\bibitem [{\citenamefont {Karki}\ and\ \citenamefont
  {Kiselev}(2020)}]{karki2020quantum}%
  \BibitemOpen
  \bibfield  {author} {\bibinfo {author} {\bibfnamefont {D.}~\bibnamefont
  {Karki}}\ and\ \bibinfo {author} {\bibfnamefont {M.~N.}\ \bibnamefont
  {Kiselev}},\ }\href@noop {} {\bibfield  {journal} {\bibinfo  {journal}
  {Physical Review B}\ }\textbf {\bibinfo {volume} {102}},\ \bibinfo {pages}
  {241402} (\bibinfo {year} {2020})}\BibitemShut {NoStop}%
\bibitem [{\citenamefont {van Dalum}\ \emph
  {et~al.}(2020{\natexlab{a}})\citenamefont {van Dalum}, \citenamefont
  {Mitchell},\ and\ \citenamefont {Fritz}}]{van2020wiedemann}%
  \BibitemOpen
  \bibfield  {author} {\bibinfo {author} {\bibfnamefont {G.~A.}\ \bibnamefont
  {van Dalum}}, \bibinfo {author} {\bibfnamefont {A.~K.}\ \bibnamefont
  {Mitchell}}, \ and\ \bibinfo {author} {\bibfnamefont {L.}~\bibnamefont
  {Fritz}},\ }\href@noop {} {\bibfield  {journal} {\bibinfo  {journal}
  {Physical Review B}\ }\textbf {\bibinfo {volume} {102}},\ \bibinfo {pages}
  {041111R} (\bibinfo {year} {2020}{\natexlab{a}})}\BibitemShut {NoStop}%
\bibitem [{\citenamefont {van Dalum}\ \emph
  {et~al.}(2020{\natexlab{b}})\citenamefont {van Dalum}, \citenamefont
  {Mitchell},\ and\ \citenamefont {Fritz}}]{van2020electric}%
  \BibitemOpen
  \bibfield  {author} {\bibinfo {author} {\bibfnamefont {G.}~\bibnamefont {van
  Dalum}}, \bibinfo {author} {\bibfnamefont {A.}~\bibnamefont {Mitchell}}, \
  and\ \bibinfo {author} {\bibfnamefont {L.}~\bibnamefont {Fritz}},\
  }\href@noop {} {\bibfield  {journal} {\bibinfo  {journal} {Physical Review
  B}\ }\textbf {\bibinfo {volume} {102}},\ \bibinfo {pages} {205137} (\bibinfo
  {year} {2020}{\natexlab{b}})}\BibitemShut {NoStop}%
\bibitem [{\citenamefont {Lee}\ \emph {et~al.}(2020)\citenamefont {Lee},
  \citenamefont {Han},\ and\ \citenamefont {Sim}}]{lee2020fractional}%
  \BibitemOpen
  \bibfield  {author} {\bibinfo {author} {\bibfnamefont {J.-Y.~M.}\
  \bibnamefont {Lee}}, \bibinfo {author} {\bibfnamefont {C.}~\bibnamefont
  {Han}}, \ and\ \bibinfo {author} {\bibfnamefont {H.-S.}\ \bibnamefont
  {Sim}},\ }\href@noop {} {\bibfield  {journal} {\bibinfo  {journal} {Physical
  Review Letters}\ }\textbf {\bibinfo {volume} {125}},\ \bibinfo {pages}
  {196802} (\bibinfo {year} {2020})}\BibitemShut {NoStop}%
\bibitem [{\citenamefont {Le~Hur}\ and\ \citenamefont
  {Simon}(2003)}]{le2003smearing}%
  \BibitemOpen
  \bibfield  {author} {\bibinfo {author} {\bibfnamefont {K.}~\bibnamefont
  {Le~Hur}}\ and\ \bibinfo {author} {\bibfnamefont {P.}~\bibnamefont {Simon}},\
  }\href@noop {} {\bibfield  {journal} {\bibinfo  {journal} {Physical Review
  B}\ }\textbf {\bibinfo {volume} {67}},\ \bibinfo {pages} {201308} (\bibinfo
  {year} {2003})}\BibitemShut {NoStop}%
\bibitem [{\citenamefont {Affleck}(1990)}]{affleck1990current}%
  \BibitemOpen
  \bibfield  {author} {\bibinfo {author} {\bibfnamefont {I.}~\bibnamefont
  {Affleck}},\ }\href@noop {} {\bibfield  {journal} {\bibinfo  {journal}
  {Nuclear Physics B}\ }\textbf {\bibinfo {volume} {336}},\ \bibinfo {pages}
  {517} (\bibinfo {year} {1990})}\BibitemShut {NoStop}%
\bibitem [{\citenamefont {Affleck}\ and\ \citenamefont
  {Ludwig}(1991{\natexlab{b}})}]{affleck1991critical}%
  \BibitemOpen
  \bibfield  {author} {\bibinfo {author} {\bibfnamefont {I.}~\bibnamefont
  {Affleck}}\ and\ \bibinfo {author} {\bibfnamefont {A.~W.}\ \bibnamefont
  {Ludwig}},\ }\href@noop {} {\bibfield  {journal} {\bibinfo  {journal}
  {Nuclear Physics B}\ }\textbf {\bibinfo {volume} {360}},\ \bibinfo {pages}
  {641} (\bibinfo {year} {1991}{\natexlab{b}})}\BibitemShut {NoStop}%
\bibitem [{\citenamefont {Affleck}(1995)}]{affleck1995conformal}%
  \BibitemOpen
  \bibfield  {author} {\bibinfo {author} {\bibfnamefont {I.}~\bibnamefont
  {Affleck}},\ }\href@noop {} {\bibfield  {journal} {\bibinfo  {journal} {arXiv
  preprint cond-mat/9512099}\ } (\bibinfo {year} {1995})}\BibitemShut {NoStop}%
\bibitem [{\citenamefont {Emery}\ and\ \citenamefont
  {Kivelson}(1992)}]{emery1992mapping}%
  \BibitemOpen
  \bibfield  {author} {\bibinfo {author} {\bibfnamefont {V.}~\bibnamefont
  {Emery}}\ and\ \bibinfo {author} {\bibfnamefont {S.}~\bibnamefont
  {Kivelson}},\ }\href@noop {} {\bibfield  {journal} {\bibinfo  {journal}
  {Physical Review B}\ }\textbf {\bibinfo {volume} {46}},\ \bibinfo {pages}
  {10812} (\bibinfo {year} {1992})}\BibitemShut {NoStop}%
\bibitem [{\citenamefont {Wilson}(1975)}]{wilson1975renormalization}%
  \BibitemOpen
  \bibfield  {author} {\bibinfo {author} {\bibfnamefont {K.~G.}\ \bibnamefont
  {Wilson}},\ }\href@noop {} {\bibfield  {journal} {\bibinfo  {journal}
  {Reviews of modern physics}\ }\textbf {\bibinfo {volume} {47}},\ \bibinfo
  {pages} {773} (\bibinfo {year} {1975})}\BibitemShut {NoStop}%
\bibitem [{\citenamefont {Bulla}\ \emph {et~al.}(2008)\citenamefont {Bulla},
  \citenamefont {Costi},\ and\ \citenamefont {Pruschke}}]{bulla2008numerical}%
  \BibitemOpen
  \bibfield  {author} {\bibinfo {author} {\bibfnamefont {R.}~\bibnamefont
  {Bulla}}, \bibinfo {author} {\bibfnamefont {T.~A.}\ \bibnamefont {Costi}}, \
  and\ \bibinfo {author} {\bibfnamefont {T.}~\bibnamefont {Pruschke}},\
  }\href@noop {} {\bibfield  {journal} {\bibinfo  {journal} {Reviews of Modern
  Physics}\ }\textbf {\bibinfo {volume} {80}},\ \bibinfo {pages} {395}
  (\bibinfo {year} {2008})}\BibitemShut {NoStop}%
\bibitem [{\citenamefont {Mitchell}\ \emph {et~al.}(2014)\citenamefont
  {Mitchell}, \citenamefont {Galpin}, \citenamefont {Wilson-Fletcher},
  \citenamefont {Logan},\ and\ \citenamefont
  {Bulla}}]{mitchell2014generalized}%
  \BibitemOpen
  \bibfield  {author} {\bibinfo {author} {\bibfnamefont {A.~K.}\ \bibnamefont
  {Mitchell}}, \bibinfo {author} {\bibfnamefont {M.~R.}\ \bibnamefont
  {Galpin}}, \bibinfo {author} {\bibfnamefont {S.}~\bibnamefont
  {Wilson-Fletcher}}, \bibinfo {author} {\bibfnamefont {D.~E.}\ \bibnamefont
  {Logan}}, \ and\ \bibinfo {author} {\bibfnamefont {R.}~\bibnamefont
  {Bulla}},\ }\href@noop {} {\bibfield  {journal} {\bibinfo  {journal}
  {Physical Review B}\ }\textbf {\bibinfo {volume} {89}},\ \bibinfo {pages}
  {121105} (\bibinfo {year} {2014})}\BibitemShut {NoStop}%
\bibitem [{\citenamefont {Stadler}\ \emph {et~al.}(2016)\citenamefont
  {Stadler}, \citenamefont {Mitchell}, \citenamefont {von Delft},\ and\
  \citenamefont {Weichselbaum}}]{stadler2016interleaved}%
  \BibitemOpen
  \bibfield  {author} {\bibinfo {author} {\bibfnamefont {K.}~\bibnamefont
  {Stadler}}, \bibinfo {author} {\bibfnamefont {A.}~\bibnamefont {Mitchell}},
  \bibinfo {author} {\bibfnamefont {J.}~\bibnamefont {von Delft}}, \ and\
  \bibinfo {author} {\bibfnamefont {A.}~\bibnamefont {Weichselbaum}},\
  }\href@noop {} {\bibfield  {journal} {\bibinfo  {journal} {Physical Review
  B}\ }\textbf {\bibinfo {volume} {93}},\ \bibinfo {pages} {235101} (\bibinfo
  {year} {2016})}\BibitemShut {NoStop}%
\bibitem [{\citenamefont {Anders}\ \emph {et~al.}(2004)\citenamefont {Anders},
  \citenamefont {Lebanon},\ and\ \citenamefont {Schiller}}]{anders2004coulomb}%
  \BibitemOpen
  \bibfield  {author} {\bibinfo {author} {\bibfnamefont {F.~B.}\ \bibnamefont
  {Anders}}, \bibinfo {author} {\bibfnamefont {E.}~\bibnamefont {Lebanon}}, \
  and\ \bibinfo {author} {\bibfnamefont {A.}~\bibnamefont {Schiller}},\
  }\href@noop {} {\bibfield  {journal} {\bibinfo  {journal} {Physical Review
  B}\ }\textbf {\bibinfo {volume} {70}},\ \bibinfo {pages} {201306} (\bibinfo
  {year} {2004})}\BibitemShut {NoStop}%
\bibitem [{SM()}]{SM}%
  \BibitemOpen
  \href@noop {} {}\bibinfo {note} {See Supplemental Material for (i) additional
  NRG data; (ii) details of NRG and conductance calculations; (iii) further
  information on the CFT treatment and Emery-Kivelson mapping. Additional
  references
  \cite{weichselbaum2007sum,meir1992landauer,mitchell2012two,ye1997solution}
  are contained therein.}\BibitemShut {Stop}%
\bibitem [{\citenamefont {Affleck}\ \emph {et~al.}(1992)\citenamefont
  {Affleck}, \citenamefont {Ludwig}, \citenamefont {Pang},\ and\ \citenamefont
  {Cox}}]{affleck1992relevance}%
  \BibitemOpen
  \bibfield  {author} {\bibinfo {author} {\bibfnamefont {I.}~\bibnamefont
  {Affleck}}, \bibinfo {author} {\bibfnamefont {A.~W.}\ \bibnamefont {Ludwig}},
  \bibinfo {author} {\bibfnamefont {H.-B.}\ \bibnamefont {Pang}}, \ and\
  \bibinfo {author} {\bibfnamefont {D.}~\bibnamefont {Cox}},\ }\href@noop {}
  {\bibfield  {journal} {\bibinfo  {journal} {Physical Review B}\ }\textbf
  {\bibinfo {volume} {45}},\ \bibinfo {pages} {7918} (\bibinfo {year}
  {1992})}\BibitemShut {NoStop}%
\bibitem [{\citenamefont {Le~Hur}\ \emph {et~al.}(2004)\citenamefont {Le~Hur},
  \citenamefont {Simon},\ and\ \citenamefont {Borda}}]{le2004maximized}%
  \BibitemOpen
  \bibfield  {author} {\bibinfo {author} {\bibfnamefont {K.}~\bibnamefont
  {Le~Hur}}, \bibinfo {author} {\bibfnamefont {P.}~\bibnamefont {Simon}}, \
  and\ \bibinfo {author} {\bibfnamefont {L.}~\bibnamefont {Borda}},\
  }\href@noop {} {\bibfield  {journal} {\bibinfo  {journal} {Physical Review
  B}\ }\textbf {\bibinfo {volume} {69}},\ \bibinfo {pages} {045326} (\bibinfo
  {year} {2004})}\BibitemShut {NoStop}%
\bibitem [{eq1()}]{eq17}%
  \BibitemOpen
  \href@noop {} {}\bibinfo {note} {See Eq.~17 in
  Ref.~\onlinecite{le2004maximized}.}\BibitemShut {Stop}%
\bibitem [{eq2()}]{eq20}%
  \BibitemOpen
  \href@noop {} {}\bibinfo {note} {See Eq.~20 in
  Ref.~\onlinecite{le2004maximized} with $V_\perp=Q_z=0$.}\BibitemShut {Stop}%
\bibitem [{\citenamefont {Georgi}(2018)}]{georgi2018lie}%
  \BibitemOpen
  \bibfield  {author} {\bibinfo {author} {\bibfnamefont {H.}~\bibnamefont
  {Georgi}},\ }\href@noop {} {\emph {\bibinfo {title} {Lie algebras in particle
  physics: from isospin to unified theories}}}\ (\bibinfo  {publisher} {CRC
  Press},\ \bibinfo {year} {2018})\BibitemShut {NoStop}%
\bibitem [{\citenamefont {Francesco}\ \emph {et~al.}(2012)\citenamefont
  {Francesco}, \citenamefont {Mathieu},\ and\ \citenamefont
  {S{\'e}n{\'e}chal}}]{francesco2012conformal}%
  \BibitemOpen
  \bibfield  {author} {\bibinfo {author} {\bibfnamefont {P.}~\bibnamefont
  {Francesco}}, \bibinfo {author} {\bibfnamefont {P.}~\bibnamefont {Mathieu}},
  \ and\ \bibinfo {author} {\bibfnamefont {D.}~\bibnamefont
  {S{\'e}n{\'e}chal}},\ }\href@noop {} {\emph {\bibinfo {title} {Conformal
  field theory}}}\ (\bibinfo  {publisher} {Springer Science \& Business
  Media},\ \bibinfo {year} {2012})\BibitemShut {NoStop}%
\bibitem [{\citenamefont {Pustilnik}\ \emph {et~al.}(2004)\citenamefont
  {Pustilnik}, \citenamefont {Borda}, \citenamefont {Glazman},\ and\
  \citenamefont {Von~Delft}}]{pustilnik2004quantum}%
  \BibitemOpen
  \bibfield  {author} {\bibinfo {author} {\bibfnamefont {M.}~\bibnamefont
  {Pustilnik}}, \bibinfo {author} {\bibfnamefont {L.}~\bibnamefont {Borda}},
  \bibinfo {author} {\bibfnamefont {L.}~\bibnamefont {Glazman}}, \ and\
  \bibinfo {author} {\bibfnamefont {J.}~\bibnamefont {Von~Delft}},\ }\href@noop
  {} {\bibfield  {journal} {\bibinfo  {journal} {Physical Review B}\ }\textbf
  {\bibinfo {volume} {69}},\ \bibinfo {pages} {115316} (\bibinfo {year}
  {2004})}\BibitemShut {NoStop}%
\bibitem [{\citenamefont {Sela}\ \emph {et~al.}(2011)\citenamefont {Sela},
  \citenamefont {Mitchell},\ and\ \citenamefont {Fritz}}]{sela2011exact}%
  \BibitemOpen
  \bibfield  {author} {\bibinfo {author} {\bibfnamefont {E.}~\bibnamefont
  {Sela}}, \bibinfo {author} {\bibfnamefont {A.~K.}\ \bibnamefont {Mitchell}},
  \ and\ \bibinfo {author} {\bibfnamefont {L.}~\bibnamefont {Fritz}},\
  }\href@noop {} {\bibfield  {journal} {\bibinfo  {journal} {Physical review
  letters}\ }\textbf {\bibinfo {volume} {106}},\ \bibinfo {pages} {147202}
  (\bibinfo {year} {2011})}\BibitemShut {NoStop}%
\bibitem [{\citenamefont {Mitchell}\ and\ \citenamefont
  {Sela}(2012)}]{mitchell2012universal}%
  \BibitemOpen
  \bibfield  {author} {\bibinfo {author} {\bibfnamefont {A.~K.}\ \bibnamefont
  {Mitchell}}\ and\ \bibinfo {author} {\bibfnamefont {E.}~\bibnamefont
  {Sela}},\ }\href@noop {} {\bibfield  {journal} {\bibinfo  {journal} {Physical
  Review B}\ }\textbf {\bibinfo {volume} {85}},\ \bibinfo {pages} {235127}
  (\bibinfo {year} {2012})}\BibitemShut {NoStop}%
\bibitem [{\citenamefont {Von~Delft}\ and\ \citenamefont
  {Schoeller}(1998)}]{von1998bosonization}%
  \BibitemOpen
  \bibfield  {author} {\bibinfo {author} {\bibfnamefont {J.}~\bibnamefont
  {Von~Delft}}\ and\ \bibinfo {author} {\bibfnamefont {H.}~\bibnamefont
  {Schoeller}},\ }\href@noop {} {\bibfield  {journal} {\bibinfo  {journal}
  {Annalen der Physik}\ }\textbf {\bibinfo {volume} {7}},\ \bibinfo {pages}
  {225} (\bibinfo {year} {1998})}\BibitemShut {NoStop}%
\bibitem [{\citenamefont {Liberman}\ \emph {et~al.}(2021)\citenamefont
  {Liberman}, \citenamefont {Mitchell}, \citenamefont {Affleck},\ and\
  \citenamefont {Sela}}]{inprep}%
  \BibitemOpen
  \bibfield  {author} {\bibinfo {author} {\bibfnamefont {A.}~\bibnamefont
  {Liberman}}, \bibinfo {author} {\bibfnamefont {A.~K.}\ \bibnamefont
  {Mitchell}}, \bibinfo {author} {\bibfnamefont {I.}~\bibnamefont {Affleck}}, \
  and\ \bibinfo {author} {\bibfnamefont {E.}~\bibnamefont {Sela}},\ }\href@noop
  {} {\bibfield  {journal} {\bibinfo  {journal} {arXiv preprint
  arXiv:2103.10680}\ } (\bibinfo {year} {2021})}\BibitemShut {NoStop}%
\bibitem [{\citenamefont {Weichselbaum}\ and\ \citenamefont {von
  Delft}(2007)}]{weichselbaum2007sum}%
  \BibitemOpen
  \bibfield  {author} {\bibinfo {author} {\bibfnamefont {A.}~\bibnamefont
  {Weichselbaum}}\ and\ \bibinfo {author} {\bibfnamefont {J.}~\bibnamefont {von
  Delft}},\ }\href@noop {} {\bibfield  {journal} {\bibinfo  {journal} {Physical
  review letters}\ }\textbf {\bibinfo {volume} {99}},\ \bibinfo {pages}
  {076402} (\bibinfo {year} {2007})}\BibitemShut {NoStop}%
\bibitem [{\citenamefont {Meir}\ and\ \citenamefont
  {Wingreen}(1992)}]{meir1992landauer}%
  \BibitemOpen
  \bibfield  {author} {\bibinfo {author} {\bibfnamefont {Y.}~\bibnamefont
  {Meir}}\ and\ \bibinfo {author} {\bibfnamefont {N.~S.}\ \bibnamefont
  {Wingreen}},\ }\href@noop {} {\bibfield  {journal} {\bibinfo  {journal}
  {Physical review letters}\ }\textbf {\bibinfo {volume} {68}},\ \bibinfo
  {pages} {2512} (\bibinfo {year} {1992})}\BibitemShut {NoStop}%
\bibitem [{\citenamefont {Mitchell}\ \emph {et~al.}(2012)\citenamefont
  {Mitchell}, \citenamefont {Sela},\ and\ \citenamefont
  {Logan}}]{mitchell2012two}%
  \BibitemOpen
  \bibfield  {author} {\bibinfo {author} {\bibfnamefont {A.~K.}\ \bibnamefont
  {Mitchell}}, \bibinfo {author} {\bibfnamefont {E.}~\bibnamefont {Sela}}, \
  and\ \bibinfo {author} {\bibfnamefont {D.~E.}\ \bibnamefont {Logan}},\
  }\href@noop {} {\bibfield  {journal} {\bibinfo  {journal} {Physical review
  letters}\ }\textbf {\bibinfo {volume} {108}},\ \bibinfo {pages} {086405}
  (\bibinfo {year} {2012})}\BibitemShut {NoStop}%
\bibitem [{\citenamefont {Ye}(1997)}]{ye1997solution}%
  \BibitemOpen
  \bibfield  {author} {\bibinfo {author} {\bibfnamefont {J.}~\bibnamefont
  {Ye}},\ }\href@noop {} {\bibfield  {journal} {\bibinfo  {journal} {Physical
  Review B}\ }\textbf {\bibinfo {volume} {56}},\ \bibinfo {pages} {R489}
  (\bibinfo {year} {1997})}\BibitemShut {NoStop}%
\end{thebibliography}
\end{document}


\renewcommand{\theequation}{S\arabic{equation}}
\renewcommand{\thefigure}{S\arabic{figure}}
\renewcommand{\thesection}{S-\arabic{section}} 
\renewcommand{\thesubsection}{S-\arabic{section}.\arabic{subsection}}
\newcommand{\be}{\begin{equation}}
\newcommand{\ee}{\end{equation}}
\newcommand{\bea}{\begin{eqnarray}}
\newcommand{\eea}{\end{eqnarray}}


\title{Supplemental Material: SO(5) non-Fermi liquid in a Coulomb box device}

\author{Andrew K. Mitchell}
\affiliation{School of Physics, University College Dublin, Belfield, Dublin 4, Ireland}
\author{Alon Liberman}
\author{Eran Sela}
\affiliation{School of Physics and Astronomy,
Tel Aviv University, Tel Aviv 6997801, Israel}
\author{Ian Affleck}
\affiliation{Department of Physics and Astronomy and Stewart Blusson Quantum Matter Institute, 
University of British Columbia, Vancouver, B.C., Canada, V6T 1Z1}

\maketitle


\section{Supplementary data}
\begin{figure}[t]
\includegraphics[width=7.2cm]{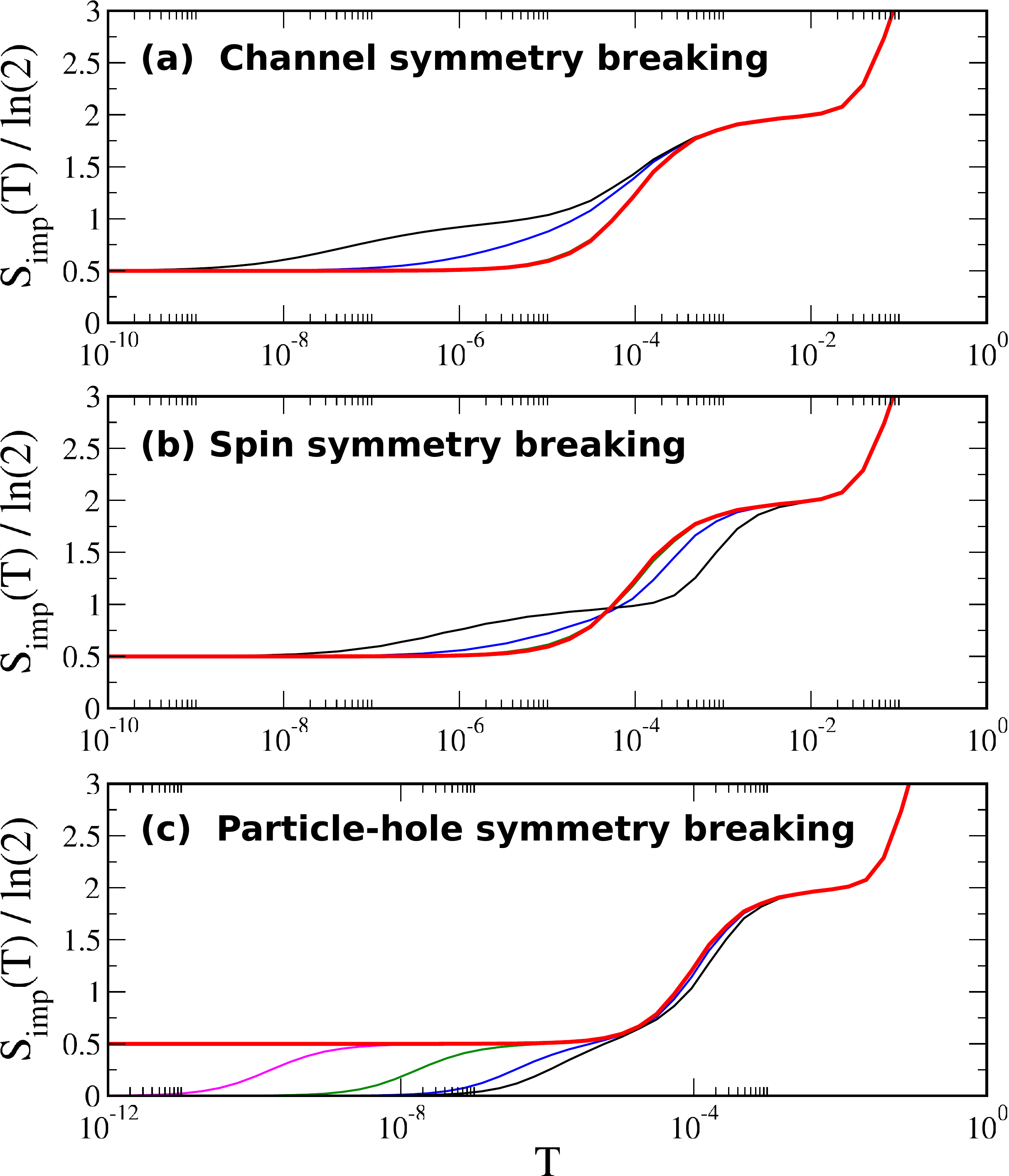}
  \caption{Dot contribution to the total entropy $S_{\rm imp}(T)$ vs $T$ obtained by NRG at and in the vicinity of the SO(5) symmetric point. The SO(5) point for $n_{\rm B}=\tfrac{1}{2}$, $\eta=0$, $U_d=0.3$, $E_C=0.1$, $t_{\rm L}=0.085$, and $t_{\rm B}\simeq \xi t_{\rm L}=0.1$ is shown as the bold red lines in each panel: the entropy flows from $\ln(4)$ for free impurity spin and flavor degrees of freedom, to $\tfrac{1}{2}\ln(2)$ characteristic of the NFL SO(5) fixed point, on the scale of the Kondo temperature $T_{\rm K}^{\rm SO(5)}\sim 10^{-4}$. (a) Channel symmetry breaking, parametrized by $\lambda=\xi^2t_{\rm L}^2/t_{\rm B}^2=1,10,33$ for the red, blue, and black lines respectively, with fixed $t_{\rm B}^2(1+\lambda)=0.02$. NFL physics is seen to be robust to this perturbation, with $S_{\rm imp}(0)=\tfrac{1}{2}\ln(2)$ in all cases. Strong channel asymmetry generates a two-stage Kondo effect with first-stage single-channel screening on the scale of $T_{\rm K}^s$ of the spin by the more strongly coupled channel, followed by two-channel overscreening of the flavor degrees of freedom on the scale of $T_{\rm K}^f$. (b) Spin symmetry breaking, due to a dot magnetic field $B/T_{\rm K}^{\rm SO(5)}=0,1,3.3,10$ for the red, green, blue and black lines respectively. For magnetic fields $B/T_{\rm K}^{\rm SO(5)}\gg 1$, the dot spin is effectively frozen, suppressing spin-Kondo screening. The free flavor degree of freedom (giving a $\ln(2)$ entropy contribution) is then two-channel overscreened below $T_{\rm K}^f$. (c) Particle-hole symmetry breaking, due to $\eta = \epsilon_d+\tfrac{1}{2}U_d = 0, 10^{-4}, 10^{-3}, 10^{-2.5}, 10^{-2}$ for the red, pink, green, blue, and black lines respectively. Ph asymmetry is seen to destabilize the NFL fixed point, generating a second Fermi liquid (FL) scale $T^*\sim \eta^2$, below which the residual impurity entropy is quenched, $S_{\rm imp}(0)=0$, at an SU(4) symmetric fixed point. With $T^*\ll T_{\rm K}^{\rm SO(5)}$, this FL crossover from SO(5) to SU(4) Kondo takes a universal form.
  }
  \label{SM:SO(5)}
\end{figure}

Figure \ref{SM:SO(5)} shows the impurity contribution to the total entropy $S_{\rm imp}(T)$ vs temperature $T$ obtained by NRG at and in the vicinity of the SO(5) symmetric point. (a) channel asymmetry, $t_{\rm B}/t_{\rm L}\ne \xi$; (b) spin asymmetry due to a dot magnetic field, $B\ne 0$; (c) ph asymmetry due to $\eta \ne 0$. At $n_{\rm B}=\tfrac{1}{2}$, NFL physics is robust to channel and spin asymmetry, but unstable to ph symmetry breaking.



\section{Details of NRG calculations}
Following Ref.~\onlinecite{anders2004coulomb} we implement with NRG the model $H_{\rm ALS} = \sum_{\alpha={\rm L,B}} (H_0^{\alpha}+H_{\rm hyb}^{\alpha}) +H_{\rm B} + H_{\rm d}$, where,
\begin{align}
 H_0^{\alpha} &=\sum_{k,\sigma}\epsilon_{\alpha k}^{\phantom{\dagger}}  c_{\alpha k\sigma}^{\dagger}c_{\alpha k\sigma}^{\phantom{\dagger}} \;, \\
H_{\rm B} &=E_C\left ( \hat{\mathcal{T}}^z-n_{\rm B} \right )^2 \;,\\
H_{\rm d} &= \sum_{\sigma} \epsilon_{\rm d}d_{\sigma}^{\dagger} d_{\sigma}^{\phantom{\dagger}} + U_{\rm d} d_{\uparrow}^{\dagger} d_{\uparrow}^{\phantom{\dagger}} d_{\downarrow}^{\dagger} d_{\downarrow}^{\phantom{\dagger}}\;,\\
H_{\rm hyb}^{\rm L} &= \sum_{k,\sigma}(t_{{\rm L} k}d_{\sigma}^{\dagger}c_{{\rm L} k\sigma}^{\phantom{\dagger}} + \rm H.c.) \\
H_{\rm hyb}^{\rm B} &= \sum_{k,\sigma}(t_{{\rm B} k}d_{\sigma}^{\dagger}c_{{\rm B} k\sigma}^{\phantom{\dagger}}\hat{\mathcal{T}}^{-} + \rm H.c.) \;,
\end{align}
where $\hat{\mathcal{T}}^z=\sum_{n=-m}^{m} n|n\rangle\langle n|$ and $\hat{\mathcal{T}}^{+}=\sum_{n=-m}^{m-1}|n+1\rangle\langle n|$ with $\hat{\mathcal{T}}^{-}=(\hat{\mathcal{T}}^{+})^{\dagger}$ in terms of the box charge states $|n\rangle$. In practice we take $m=2$, which yields well-converged low-temperature results for $E_C=0.1$ and $t_{\alpha}^2=\sum_k |t_{\alpha k}|^2 \sim \mathcal{O}(10^{-2})$ as used here. For simplicity we take equivalent lead and box bands, with a flat conduction electron density $\nu$ within bands of halfwidth $D\equiv 1$. We define $\eta=\epsilon_d+\tfrac{1}{2}U_d$ as before.

The free conduction electron bands are discretized logarithmically using $\Lambda=2.5$, and $N_s\sim 8000$ states are retained at every step of the iterative diagonalization procedure.\cite{wilson1975renormalization,*bulla2008numerical} We utilized the `interleaved NRG' method on the generalized Wilson chain,\cite{mitchell2014generalized,*stadler2016interleaved} exploiting conserved total charge and spin projection. Dynamical quantities are calculated using the full density matrix NRG method.\cite{weichselbaum2007sum}


\clearpage
\section{Conductance}
In the `split-lead' geometry utilized in the experiments of Refs.~\onlinecite{potok2007observation,keller2015universal} and depicted in Fig.~1 of the main paper, the free lead Hamiltonian is decomposed as $H_0^{\rm L}=\sum_{\gamma={\rm Ls,Ld}}\sum_{k,\sigma}\epsilon_{{\rm L} k}^{\phantom{\dagger}}  c_{\gamma k\sigma}^{\dagger}c_{\gamma k\sigma}^{\phantom{\dagger}}$ into source and drain components, while $H_{\rm hyb}^{\rm L} = \sum_{\gamma={\rm Ls,Ld}}\sum_{k,\sigma}(t_{\gamma k}d_{\sigma}^{\dagger}c_{\gamma k\sigma}^{\phantom{\dagger}} + \rm H.c.)$, and with $t_{\gamma}^2 = \sum_k |t_{\gamma k}|^2$.

A voltage bias, $V_{\rm bias}$, is applied symmetrically to the leads, $H_{\rm bias}=\tfrac{1}{2}eV_{\rm bias}(\hat{N}_{\rm Ls}-\hat{N}_{\rm Ld})$ with $\hat{N}_{\gamma}=\sum_{k,\sigma}c_{\gamma k\sigma}^{\dagger}c_{\gamma k\sigma}^{\phantom{\dagger}}$, and we measure the current into the drain lead, $I_{\rm Ld}=\langle \tfrac{d}{dt} \hat{N}_{\rm Ld}\rangle$. The linear-response differential conductance,
\begin{equation}
    G=\lim_{V_{\rm bias}\to 0} ~\frac{d I_{\rm Ld}}{dV_{\rm bias}} \;,
\end{equation}
is related to the equilibrium dot Green's function $G_{dd}(\omega,T)\equiv \langle \langle d_{\sigma} ; d_{\sigma}^{\dagger}\rangle\rangle$ via the Meir-Wingreen formula,\cite{meir1992landauer}
\begin{equation}
G(T) = G_0\int_{-\infty}^{\infty} d\omega ~\frac{df(\omega)}{d\omega} \left[ \pi \nu t_{\rm L}^2 ~{\rm Im}~ G_{dd}(\omega,T)\right ] \;, 
\end{equation}
where $f(\omega)$ is the Fermi function, and $G_0=2e^2/h\times 4t_{\rm Ls}^2 t_{\rm Ld}^2/t_{\rm L}^4$. We obtain the dot Green's function $G_{dd}(\omega,T)$ as an entire function of (real) frequency $\omega$ at a given temperature $T$ using NRG as above.

We find $\pi \nu t_{\rm L}^2~ {\rm Im}~G_{dd}(0,0)=\tfrac{1}{2}$, implying $G(0)=\tfrac{1}{2}G_0$, everywhere along the NFL line. Deep in the Coulomb blockade regime of the box where the standard spin-2CK effect is observed, $G(T)-G(0)\sim -\sqrt{T/T_{\rm K}}G_0$. At $n_{\rm B}=\tfrac{1}{2}$ and $\eta=0$ where spin and flavor fluctuations both play an important role in driving the Kondo effect, we obtain $G(T)-G(0)\sim \pm \sqrt{T/T_{\rm K}}G_0$, with the $+$ or $-$ sign, respectively, for channel asymmetry $t_{\rm B}>\xi t_{\rm L}$ or $t_{\rm B}<\xi t_{\rm L}$. This is due to a two-stage Kondo effect in the strongly channel asymmetric case. However, when precisely at the SO(5) point, where $t_{\rm B}=\xi t_{\rm L}$, the square-root behavior vanishes, 
leaving a leading temperature dependence $G(T)-G(0)\sim +(T/T_{\rm K})^{3/2} G_0$ (see Fig.~2 of the main paper). This is reminiscent of the behavior of conductance in the two-impurity Kondo model.\cite{mitchell2012two} 

At the SO(5) point, the full conductance lineshape is a universal function of $T/T_K$, and is entirely characteristic of the novel SO(5) Kondo effect. This should be measurable in experiment (note also that the Kondo temperature itself is enhanced at the SO(5) point).


\section{CFT treatment of the\\SO(5) fixed point}

In this appendix we derive the finite size spectrum at the SO(5) symmetric fixed point using CFT. For a review of CFT methods for Kondo problems see Ref.~\onlinecite{affleck1995conformal}.

We employ a U(1)$_{c} \times Z_2 \times$ SO(5)$_1$  conformal embedding. Here, U(1) stands for the charge sector, whose primary fields are labelled by integer charges $Q$  and have scaling dimension $\frac{Q^2}{8}$; $Z_2$ is the Ising model consisting of  three primary fields: the identity field of scaling dimension $0$, the spin field $\sigma$ of scaling dimension $\frac{1}{16}$, and the fermion field $\epsilon$ with scaling dimension 1/2, which satisfy the fusion rules $\sigma \times \sigma = 1 + \epsilon$, $\sigma \times \epsilon = \sigma$; primary fields of the SO(5)$_1$ theory are labelled by  SO(5) representations, consisting of~\cite{francesco2012conformal} the singlet  ${\bf{(1)}}$ of scaling dimension 0, the spinor  ${\bf{(4)}}$ of scaling dimension $\frac{5}{16}$ and the vector ${\bf{(5)}}$ of scaling dimension $1/2$. The SO(5)$_1$ theory can also be thought of as 5 Ising models, and its fusion rules  follow by  identifying the fields ${\bf{(1)}},{\bf{(4)}},{\bf{(5)}}$ with $1,\sigma,\epsilon$. We note that other representations of SO(5) appear as descendants of the above primary fields, for example $J^A=c^\dagger T^A c$ which transforms under the $10$ dimensional representation, similar to the impurity spin and flavour degrees of freedom, is a (Kac-Moody) descendent of the singlet ${\bf{(1)}}$. 

\begin{table}[h!]
	\begin{center}
		\caption{Finite size spectrum.}
		\label{tab:table1}
		\begin{tabular}{c|c|c||c ||| c|c|c||c|} 
			\multicolumn{4}{c|||}{Free spectrum} &			\multicolumn{4}{c|}{Fusion with ${\bf{(4)}}$} \\
			$Q$ & Ising & SO(5) & E & 	$Q$ & Ising & SO(5) & $E-\frac{3}{16}$\\
			\hline
			0 & 1 &  ${\bf{(1)}}$ & 0  & 0 & 1 &  ${\bf{(4)}}$ & $\frac{1}{8}$
			\\
			$\pm 1$ & $\sigma$ &  ${\bf{(4)}}$ & $\frac{1}{2}$  & 	$\pm 1$ & $\sigma$ &  ${\bf{(1)}}$ & $0$
			\\
			&  &   &  & 	$\pm 1$ & $\sigma$ &  ${\bf{(5)}}$ & $\frac{1}{2}$
			\\
			$\pm 2$ & $\epsilon$ &  ${\bf{(1)}}$ & 1  & $\pm 2$ & $\epsilon$ &  ${\bf{(4)}}$ & 	\\
			$\pm 2$ & 1 &  ${\bf{(5)}}$ & 1  & 	$\pm 2$ & 1 &  ${\bf{(4)}}$ & $\frac{5}{8}$
			\\
			$0$ & $\epsilon$ &  ${\bf{(5)}}$ & 1  & 	$0$ & $\epsilon$ &  ${\bf{(4)}}$ & $\frac{5}{8}$
		\end{tabular}
	\end{center}
\end{table}

The free fermion spectrum can  be recovered by the gluing conditions shown in the left hand side part in Table \ref{tab:table1}. [As reviewed in Ref.~\onlinecite{affleck1995conformal}, energy levels are obtained by adding up scaling dimensions in units of $2 \pi v_F/L$, where $L$ is the length of the effective 1D system and $v_F$ the Fermi velocity.] We obtain the strong coupling fixed point by fusion with the spinor field ${\bf{(4)}}$ of SO(5)$_1$. Physically, the impurity is absorbed into the electrons and changes their boundary condition. The resulting finite size spectrumlisted in  the right side of Table \ref{tab:table1}, is fully consistent with our NRG observations. The first  energies and degeneracies are given by 
\be
\label{FSS2CK}
(E, \#) = (0,2) ; (1/8,4) ; (1/2,10) ; (5/8,12) ; (1,26)\dots
\ee

\section{Emery-Kivelson Hamiltonian \\of spin-flavour model}
Emery-Kivelson's (EK) bosonization and refermionization technique, originally used to solve the 2CK problem~\cite{emery1992mapping}, allows to express the Hamiltonian in a free fermion form. In this appendix we derive the EK form of the Hamiltonian of our spin-flavour Kondo model Eq.~5 and its various perturbations. We will outline the rigorous treatment of the s-2CK case~\cite{von1998bosonization} and highlight the required additions involving the flavour impurity degree of freedom. (A similar derivation for the spin-flavour Kondo model appears in Ref.~\onlinecite{ye1997solution}.)

Our starting point is the ph symmetric and channel symmetric  Hamiltonian Eq.~5 in the main text with anisotropic bare  coupling constants: (i) In the flavour sector the unperturbed model contains only a $V_z$ coupling and no a flavour-flip  ($V_\perp$), since the latter breaks ph symmetry (treated below as a perturbation). (ii) In the spin sector  we separate the SU(2) symmetric Kondo coupling $J$ into $J_z$ and $J_\perp$ components. 
(iii) Also the spin-flavour operator $Q_\perp$ in Eq.~5 in the main text is split into spin-flip part $Q_{\perp}^\perp$ and a non-spin-flip part $Q_\perp^z$.

The EK technique is applied to the hamiltonian as follows: (1) Fermionic fields  are transformed into bosonic fields $\phi_{\sigma \alpha}(x)$, according to the relation $\psi_{\sigma \alpha}(x)=F_{\sigma \alpha}e^{-i\phi_{\sigma \alpha}}$ ($F_{\sigma \alpha}$ being fermionic Klein operators). (2) The new bosonic fields are re-expressed in a basis of charge, spin, flavour and spin-flavor  degrees of freedom abbreviated as $\phi_A(x), (A={\rm{c,s,f,sf}}$). (3) A unitary operator  $U=\exp[i(S^z \phi_{\rm{s}}+T^z \phi_{\rm{f}})]$ acting \emph{both} in  spin and flavor spaces is applied on the Hamiltonian. With a particular choice of the $z$-part of the spin and flavor Kondo couplings $J_z$ and $V_z$, (i) terms of the form $\partial_{x}\phi_{\rm{s}}(0) S^z$ and $\partial_{x}\phi_{\rm{f}}(0) T^z$  effectively drop out of the Hamiltonian, and (ii)  vertex parts $e^{-i \phi_A}$ of both spin and flavour sectors cancel. To achieve both cancellations we must have specific bare couplings $J_z=V_z=v_F$.  (4) Klein operators are expressed in the basis of $A={\rm{c,s,f,sf}}$. (5) The Hamiltonian is expressed in terms of new fermionic operators: $ \psi_{\rm{sf}}(x=0)= F_{\rm{sf}}e^{-i\phi_{\rm{sf}}(0)}  ,~~d=F_{\rm{s}}^{\dagger}S^{-},~~a=F^{\dagger}_{\rm{f}}T^{-}$. These can be used to define Majorana fermion fields evaluated at $x=0$:
\bea
\label{A1}
 \chi_{A+} \equiv \frac{\psi^{\dagger}_{A}(0)+\psi_{A}(0)}{\sqrt{2}}, ~\chi_{A-}\equiv\frac{\psi^{\dagger}_{A}(0)-\psi_{A}(0)}{\sqrt{2}i}.
 \eea
Similarly we  define  local Majoranas $a_{\pm},d_{\pm}$ using the $a$ and $d$ operators for the flavour and spin impurity degrees of freedom, respectively. 
The presence of 4 local Majorana fermions, simply accounts for the $2 \log 2$ entropy of the free fermion fixed point, due to both the spin and flavour impurity  degrees of freedom $S$ and $T$. In these terms, the resulting 
Hamiltonian is
\begin{equation}
\label{eq:EKSM}
H_{EK}=H_{0}+iJ_{\perp}d_{-}\chi_{+}-2Q_{\perp}^{z}a_{-}\chi_{+}d_{-}d_{+}-2iQ_{\perp}^{\perp}d_{+}a_{-} .
\end{equation}
The  term $\propto J_{\perp}$ appears in the usual s-2CK model, which is a relevant perturbation to the free fermion fixed point $H_0$, leading to the absorption of the local Majorana $d_-$ into the Majorana field $\chi_+$.
The last term $\propto Q_{\perp}^{\perp}$  gaps out the sector spanned by $d_{+}$ and $a_{-}$, meaning that we can exchange every appearance of the product $id_{+}a_{-}$ by its expectation value. This way, the term $\propto Q_{\perp}^z$ becomes the same as $J_{\perp}$, merely renormalizing it. Eq.~7 in the main text follows (with simplified notation $Q_\perp^\perp \to Q_\perp$). 

Symmetry breaking perturbations include channel asymmetry, magnetic field, and ph symmetry breaking. Channel asymmetry and magnetic field have the same EK form as in the s-2CK model, as discussed extensively in the literature~\cite{mitchell2012universal}. We focus on ph breaking terms which are unique to the spin-flavour model.


Moving away from the point $\eta=0$, $n_{B}=\frac{1}{2}$  in the phase diagram, ph symmetry is broken. As a result, the terms $\propto V_\perp$ and $Q_z$ (missing by symmetry from Eq.~5 in the main text, and featuring in $\delta H_{\rm{ph}}$), appear. Focusing on $V_\perp T^+ c^\dagger \tau^- c+h.c.$, and performing the EK transformation, we obtain the relevant operator  $\delta H_{\rm{ph}} = iV_{\perp}a_{+}\chi_{-}$. The operator $Q_z$ generates the same relevant operator.


\vfill
%
